\documentclass[12pt]{article}

\usepackage{booktabs}
\usepackage{amsmath}
\usepackage{amssymb}
\usepackage{hhline}
\usepackage[scale={.75,.75}]{geometry}
\usepackage{url}
\usepackage{caption}
\usepackage[numbers, sort&compress]{natbib}
\usepackage{epsfig}
\usepackage{multirow}

\newcommand{\GeV}{\ \text{GeV}}
\newcommand{\TeV}{\ \text{TeV}}
\newcommand{\kpc}{\ \text{kpc}}
\newcommand{\keV}{\ \text{keV}}

\newcommand{\s}{\ \text{s}}
\newcommand{\cm}{\ \text{cm}}
\newcommand{\fex}{\textit{e.g.}~}
\newcommand{\fatc}{{\bf c}}
\newcommand{\fatmu}{{\boldsymbol \mu}}

\usepackage{color}
\newcommand{\MP}{M_{\text{P}}}
\newcommand{\GF}{G_{\text{F}}}
\newcommand{\mG}{m_{3/2}}
\newcommand{\mN}{m_{\chi_1^0}}
\newcommand{\mstau}{m_{\tilde \tau_1}}

\begin{document}
\date{\mbox{ }}

\title{ 
{\normalsize     
DESY 10-237\hfill\mbox{}\\
MPP-2010-166\hfill\mbox{}\\}
\vspace{1cm}
\bf \boldmath 
Hunting Dark Matter Gamma-Ray~Lines with the Fermi LAT\\[8mm]}

\author{Gilles Vertongen$^{1}$, Christoph~Weniger$^{2}$ \\[8mm]
{\normalsize \it $^1$ Deutsches Elektronen-Synchrotron (DESY),
 \it Notkestrasse 85, 22603 Hamburg, Germany}\\
{\normalsize \it$^2$ Max-Planck-Institut f\"ur Physik, F\"ohringer Ring 6,
80805 M\"unchen, Germany}
}
\maketitle

\thispagestyle{empty}

\begin{abstract}
  \noindent
  Monochromatic photons could be produced in the annihilation or decay of dark
  matter particles.  At high energies, the search for such line features in
  the cosmic gamma-ray spectrum is essentially background free because
  plausible astrophysical processes are not expected to produce such a signal.
  The observation of a gamma-ray line would hence be a `smoking-gun' signature
  for dark matter, making the search for such signals particularly attractive.
  Among the different dark matter models predicting gamma-ray lines, the local
  supersymmetric extension of the standard model with small $R$-parity
  violation and gravitino LSP is of particular interest because it provides a
  framework where primordial nucleosynthesis, gravitino dark matter and
  thermal leptogenesis are naturally consistent.  Using the two-years Fermi
  LAT data, we present a dedicated search for gamma-ray lines coming from dark
  matter annihilation or decay in the Galactic halo.  Taking into account the
  full detector response, and using a binned profile likelihood method, we
  search for significant line features in the energy spectrum of the diffuse
  flux observed in different regions of the sky. No evidence for a line signal
  at the $5\sigma$ level is found for photon energies between 1 and 300 GeV,
  and conservative bounds on dark matter decay rates and annihilation cross
  sections are presented. Implications for gravitino dark matter in presence
  of small $R$-parity violation are discussed, as well as the impact of our
  results on the prospect for seeing long-lived neutralinos or staus at the
  LHC.
\end{abstract}

\newpage

\section{Introduction}
Gravitational evidences for dark matter exist from galactic to cosmological
scales, but its particle nature still remains unknown~\cite{Jungman:1995df,
Bergstrom:2000pn, Bertone:2004pz}. The most popular dark matter candidate, a
weakly interacting massive particle (WIMP), could in principle be detected
directly through its scattering on atomic nuclei, indirectly by observing
traces of its annihilation products in the cosmic-ray fluxes, or by its direct
production at colliders.  Since the stability of dark matter is by no means
established, there is also the possibility to observe cosmic-ray contributions
stemming from the decay of dark matter particles. However, despite lots of
efforts, no unambiguous non-gravitational dark matter signal has been found so
far.\medskip

Indirect searches for dark matter annihilation or decay products in the cosmic
gamma-ray, anti-particle and neutrino fluxes are limited not only by the
instrument sensitivities, but also by our understanding of the astrophysical
foregrounds. One of the few dark matter signatures that would unambiguously
stand out of the astrophysical foreground are the monochromatic photons
potentially produced in two-body annihilation or decay of dark matter. They
would appear as a line feature in the otherwise continuous gamma-ray energy
spectrum~\cite{Jungman:1995df, Bergstrom:2000pn, Bertone:2004pz,
Srednicki:1985sf, Rudaz:1986db, Bergstrom:1988fp}. The observation of such a
gamma-ray line would thus be of paramount interest for the understanding of
dark matter in the Universe.

The production of gamma-ray lines is expected in many theoretical dark matter
models. The corresponding branching ratios are often very small, but this
could be compensated by the high experimental sensitivity to such signals.  In
the popular scenario where dark matter is the lightest neutralino $\chi_1^0$
of the minimal supersymmetric standard model (MSSM)~\cite{Martin:1997ns},
gamma-ray lines from $\chi_1^0\chi_1^0\to\gamma\gamma$ or
$\chi_1^0\chi_1^0\to\gamma Z^0$ annihilations are only produced at the
one-loop level~\cite{Rudaz:1990rt, Bergstrom:1997fj, Bergstrom:1997fh,
Bern:1997ng, Ullio:1997ke, Scott:2009jn, Ripken:2010ja}, and they would be
mostly out of reach for current experiments.  However, their production can be
enhanced  in non-minimal variations of the MSSM~\cite{Profumo:2008yg,
Ferrer:2006hy}. Alternatively, a large number of viable WIMP models exist,
some of them predicting a sizable annihilation into monochromatic photons,
\fex singlet dark matter~\cite{Profumo:2010kp}, hidden U(1) dark
matter~\cite{Dudas:2009uq, Mambrini:2009ad, Mambrini:2010yp}, effective dark
matter scenarios~\cite{Goodman:2010qn}, scenarios with strong annihilation
into the higgs boson and a photon~\cite{Jackson:2009kg}, inert higgs doublet
dark matter~\cite{Gustafsson:2007pc}, or Kaluza-Klein dark matter in models
with universal extra dimensions~\cite{Bertone:2009cb, Bertone:2010fn}.
Additionally, prominent spectral features in the gamma-ray flux can be
generated by the internal Bremsstrahlung or final state radiation potentially
produced in dark matter annihilation~\cite{Bergstrom:2005ss, Birkedal:2005ep,
Bringmann:2007nk, Bergstrom:2004cy}.

An interesting example of decaying dark matter is the gravitino $\psi_{3/2}$,
which appears in locally supersymmetric extensions of the Standard Model. In
scenarios where $R$-parity and the lepton number are slightly violated and the
gravitino is the lightest superparticle (LSP), thermal leptogenesis, gravitino
dark matter and primordial nucleosynthesis are naturally
consistent~\cite{Buchmuller:2007ui}. Within this framework, the gravitino is
no longer stable and decays on cosmological time
scales~\cite{Takayama:2000uz}, thus rendering the imprints of its decay
potentially observable in the cosmic-ray fluxes~\cite{Bertone:2007aw,
Ibarra:2007wg, Lola:2007rw, Ibarra:2008qg, Ishiwata:2008cu, Ishiwata:2008tp,
Covi:2008jy, Buchmuller:2009xv, Choi:2009ng, Bomark:2009zm, Choi:2010xn,
Choi:2010jt}. Interestingly, the line feature generated by the two-body decay
into neutrinos and photons, $\psi_{3/2}\to\gamma\nu$, is prominent for a wide
range of gravitino masses~\cite{Ibarra:2007wg}, suggesting that the first
observation of gravitino dark matter could happen via gamma-ray lines.

In general, dark matter lifetimes of the order of $10^{26}$--$10^{29}$s, which
is in the ballpark of what is accessible by cosmic-ray experiments in the
GeV--TeV regime, are obtained when the symmetry responsible for the dark
matter stability is violated by dimension six operators generated close to the
grand unification scale~\cite{Eichler:1989br, Arvanitaki:2009yb}.  A typical
example, which has a large branching fraction into monochromatic photons, is
the hidden SU(2) vector dark matter~\cite{Arina:2009uq}.  Note that even if
dark matter only decays into charged leptons at tree level, like in the models
designed to explain the PAMELA/Fermi~LAT~$e^\pm$
measurements~\cite{Adriani:2008zr, Abdo:2009zk, Ackermann:2010ij}, gamma-ray
lines produced at the one-loop level could potentially be
observable~\cite{Garny:2010eg}.\medskip

A dedicated search for Galactic gamma-ray lines with energies between 30 and
200 GeV using the Fermi Large Area Telescope (LAT) data~\cite{Atwood:2009ez}
has been presented in Ref.~\cite{Abdo:2010nc}, where the all-sky averaged
diffuse gamma-ray flux measured between 7~Aug 2008 and 21~Jul 2009 was
considered. Searching for monochromatic features in the energy spectrum, their
null result enable them to put constraints on the dark matter annihilation
cross section and decay rate into gamma-ray lines. A similar analysis using
the EGRET\footnote{See
\url{http://heasarc.gsfc.nasa.gov/docs/cgro/cgro/egret.html}.} data has been
performed for gamma-ray line energies between 0.1 and 10 GeV in
Ref.~\cite{Pullen:2006sy}.

The extra-galactic gamma-ray background (EGBG), as derived from the Fermi LAT
data in Ref.~\cite{Abdo:2010nz}, has also been used in Ref.~\cite{Abdo:2010dk}
to put constraints on gamma-ray lines from cosmological dark matter
annihilation~\cite{Bergstrom:2001jj, Ullio:2002pj}.  In that case, the
redshifted gamma-ray lines lead to less pronounced features in the energy
spectrum than the Galactic ones. The lack of spectral features in the EGBG was
used to put limits on the corresponding annihilation cross section. Further
gamma-line searches, coming from different observations, and ranging from
energies of $10\keV$ up to $10\TeV$, were presented for annihilation in
Ref.~\cite{Mack:2008wu} and for decay in Ref.~\cite{Yuksel:2007dr}. General
strategies for the detection of gamma-ray lines from dark matter annihilation
with Fermi LAT and future experiments were discussed in
Refs.~\cite{Serpico:2008ga, Serpico:2009vz, Tang:2010tp}.\medskip

The purpose of this paper two-fold: First, we will update and extend the
gamma-ray line analysis of the Fermi LAT data presented in
Ref.~\cite{Abdo:2010nc} to a larger energy range of $1$--$300\GeV$, searching
for significant line signals that might come from dark matter annihilating or
decaying in the Galactic dark matter halo, and derive conservative
constraints. Although highly relevant for the gravitino dark matter scenario,
no dedicated search for gamma-ray lines from decaying dark matter in the whole
$\mathcal{O}($GeV--TeV$)$ energy regime has been performed so far.
Furthermore, the large statistics of the Fermi LAT data allows us to further
probe the low energy regime, \textit{i.e.}~energies from 1 to 10 GeV, and thus
strengthening the previous EGRET bounds on annihilating dark
matter~\cite{Pullen:2006sy}.  Such energies are of particular interest in the
view of the recent hints from direct dark matter detection reported by the
CDMS~\cite{Ahmed:2009zw}, CoGeNT~\cite{Aalseth:2010vx} and
DAMA~\cite{Bernabei:2008yi, Bernabei:2010mq} collaborations.  Second, we will
discuss the impact of our results on decaying gravitino dark matter. Covering
previously unexplored values of the gravitino mass, we will use our limits on
gamma-ray lines to constrain the size of $R$-parity violation, which can be
translated into lower limits on the decay lengths of neutralino or stau
NLSPs.\medskip

The paper is organized as follows: In section 2, we present our analysis of
the Fermi LAT data. After briefly discussing the gamma-ray flux produced in
the dark matter annihilation or decay, we present the strategy for data
extraction and line searches in section 2.1. Section 2.2 is devoted to a
presentation of the resulting bounds on annihilation cross section and decay
rate, whereas in section 2.3 we discuss details and caveats of our analysis.
In section 3, we apply our results to gravitino dark matter and discuss
prospects for NLSP observations at particle colliders. We then conclude in
section~4.

\section{Gamma-ray line constraints from Fermi LAT}
Our line search is based on the measurements of the cosmic gamma-ray flux
performed since August 2008 by the Large Area Telescope (LAT), the main
instrument on the Fermi Gamma-ray Space Telescope. The LAT is a pair
conversion detector, designed to observe gamma rays with energies ranging from
30 MeV to more than 300 GeV with an energy resolution of about $10\%$. It has
a large field-of-view that spans around $20\%$ of the sky, allowing an
effective measurement of the all-sky diffuse fluxes. From the four event
classes presented in the available public data, we only consider the recently
released `DataClean' event class, which provides the best rejection of charged
cosmic rays. This is especially important when analyzing the diffuse gamma-ray
flux at high energies.\medskip 

Gamma-ray lines may find their origin in the decay or annihilation of dark
matter particles $\psi$ in the Galactic halo. The observed gamma-ray flux at
Earth is given by a line-of-sight integral depending linearly on the dark
matter distribution $\rho_\text{dm}(r)$ when considering dark matter decays
\begin{equation}
  \frac{dJ_\gamma}{dE}(\xi) =
  \frac{\Gamma_{\psi\to\gamma\nu}}{4\pi \,m_\psi}
  \,\delta\left(E-\frac{m_\psi}{2}\right) \underbrace{\int_\text{l.o.s.} ds
  \,\rho_\text{dm}(r)}_{\equiv J_\rho^\text{dec.}}\;,
  \label{eqn:fluxDDM}
\end{equation}
while in the annihilation case, it depends quadratically on
$\rho_\text{dm}(r)$
\begin{equation}
  \frac{dJ_\gamma}{dE}(\xi) = 2\,\frac{\langle \sigma
  v\rangle_{\psi\psi\to\gamma\gamma}}{8\pi \,m_\psi^2}
  \,\delta\left(E-m_\psi\right) \underbrace{\int_\text{l.o.s.} ds
  \,\rho_\text{dm}^2(r)}_{\equiv J_\rho^\text{ann.}}\;.
  \label{eqn:fluxADM}
\end{equation}
Here, $m_\psi$ is the dark matter mass, while $\Gamma_{\psi\to\gamma\nu}$ and
$\langle \sigma v\rangle_{\psi\psi\to\gamma\gamma}$ denote the dark matter
decay width and annihilation cross section into gamma-ray lines, respectively.
The coordinate $s$ runs along the line of sight, in a direction spanning an
angle $\xi$ with respect to the Galactic center direction. The distance to the
Galactic center $r$ is related to the distance to the Sun $s$ through
$r(s,\xi) = \sqrt{(r_0-s\cos\xi)^2 + (s\sin\xi)^2}$, where $r_0=8.5\kpc$ is
the distance of the Sun to the Galactic center~\cite{Ghez:2008ms,
Gillessen:2008qv}.\medskip

We consider the Navarro-Frenk-White (NFW) profile~\cite{Navarro:1996gj,
Abdo:2010nc} 
\begin{align}
  \rho_{\text{dm}}(r) = \rho_0 \left[ \frac{r}{r_0} \right]^{-\gamma} \left[
  \frac{1+(r_0/r_s)^\alpha}{1+(r/r_s)^\alpha}
  \right]^{\frac{\beta-\gamma}{\alpha}}\;,
\end{align}
which is defined by $(\alpha, \beta, \gamma) = (1,3,1)$ and $r_s=20\kpc$, as a
reference for the dark matter distribution. For the sake of comparison, we
also consider the isothermal profile with $(\alpha, \beta, \gamma) = (2,2,0)$
and $r_s=3.5\kpc$, as well as the Einasto profile~\cite{Navarro:2003ew,
Springel:2008cc, Pieri:2009je}
\begin{align}
  \rho_{\text{dm}}(r) = \rho_0 \exp \left[-\frac{2}{\alpha}\left( 
  \frac{r^\alpha - r_0^\alpha}{r_s^\alpha}  \right)  \right] \qquad
  \text{with} \quad \alpha = 0.17\;,\quad r_s=20\kpc\;,
\end{align}
which is favored by the latest $N$-body simulations.\footnote{However, see
Ref.~\cite{Salucci:2007tm} for observational arguments in favor of cored dark
matter profiles.} All profiles are normalized to
$\rho_\text{dm}(r=r_0)=0.4\GeV\cm^{-3}$ at Sun's
position~\cite{Salucci:2010qr, Catena:2009mf}. The resulting line-of-sight
integrals $J_\rho^\text{dec.}$ and $J_\rho^\text{ann.}$, integrated over our
target regions (see below), are summarized in Tab.~\ref{tab:regions} for the
NFW profile.  Note that substructures in the Galactic dark matter distribution
can change the angular profile and magnitude of the gamma-ray emission (see
\fex Refs.~\cite{Pieri:2009je, Tang:2010tp}). If this is the case, our results
can straightforwardly be generalized by adopting the corresponding values for
$J_\rho^\text{dec./ann.}$.\medskip 

\begin{table}[t]
  \centering
  \begin{tabular}{@{}cccccc@{}}
    \toprule
    Target region&\phantom{}&Geometry&\phantom{}& 
    $\int_{\Delta\Omega}d\Omega \,J_\rho^\text{dec.}$ &
    $\int_{\Delta\Omega}d\Omega \,J_\rho^\text{ann.}$\\[1mm]
    &&&&[GeV\,cm$^{-2}$\,sr]&[GeV$^2$\,cm$^{-5}$\,sr]\\
    \midrule
    Halo && $|b|\geq 10^\circ$ &
    & $2.2\times10^{23}$ & $8.3\times10^{22}$ \\[2mm]
    \multirow{2}{*}{Center} && $|l|\leq36^\circ$ and
    $5^\circ\leq|b|\leq36^\circ$ &&
    \multirow{2}{*}{$6.9\times10^{22}$} &
    \multirow{2}{*}{$9.2\times10^{22}$}\\
    && \textit{plus} $|l|\leq7^\circ$ and $|b|\leq5^\circ$&\\
    \bottomrule
  \end{tabular}
  \caption{Geometry of our target regions in terms of Galactic longitude $l$
  and latitude $b$. The line-of-sight integrals $J_\rho^\text{dec.}$ and
  $J_\rho^\text{ann.}$ integrated over the target regions are also shown,
  assuming a NFW profile.} 
  \label{tab:regions}
\end{table}

\subsection{Methods}
\label{sec:methods}
The gamma-ray events that enter our analysis are selected from the `DataClean'
event class of the Fermi LAT data measured between 4 Aug 2008 and 17 Nov
2010.\footnote{The event data, as well as the corresponding information about
the instrument response functions concerning angular and energy resolution,
are available at \url{http://fermi.gsfc.nasa.gov/ssc/data/}.} From all events
recorded by the Fermi LAT, we select energies between 100\,MeV and 500\,GeV,
and apply the zenith angle criterion $\theta<105^\circ$ in order to avoid
contamination by the Earth's Albedo.\footnote{\label{fn:gti}These selections
are made using the Fermi Science Tools v9r18p6, see
\url{http://fermi.gsfc.nasa.gov/ssc/data/analysis/software/}.  As cut in
\texttt{gtmktime} we took \texttt{DATA\_QUAL==1 \&\& LAT\_CONFIG==1 \&\&
ABS(ROCK\_ANGLE)<52}.} Below 1\,GeV, the fast variation of the effective area
as well as the presence of small and sharp jumps in the energy spectrum
(presumably related to side effects of the energy
reconstruction~\cite{Edmonds:2007zz, Fermi:caveats}) introduce spurious
effects in our line search. Furthermore, the nominal energy range of the Fermi
LAT ends at $300\GeV$~\cite{Atwood:2009ez}. As a consequence, we concentrate
on gamma-ray line energies between 1 and 300 GeV. Systematic uncertainties, as
well as limitations of the analysis method, are discussed in
section~\ref{sec:discussion} in light of the data.
\medskip

When looking for dark matter signals in gamma-ray maps that cover the whole
sky, it is critical to choose a target region which maximizes the
corresponding signal-to-noise ratio $S/N$.  The two regions that we found
giving a very good $S/N$ for decaying dark matter (`halo region') or
annihilating dark matter (`center region') are summarized in
Tab.~\ref{tab:regions}. Although optimized for the NFW profile, they also
yield good $S/N$ for the Einasto and isothermal profiles.

\begin{figure}[t]
  \begin{center}
    \includegraphics[width=0.49\linewidth]{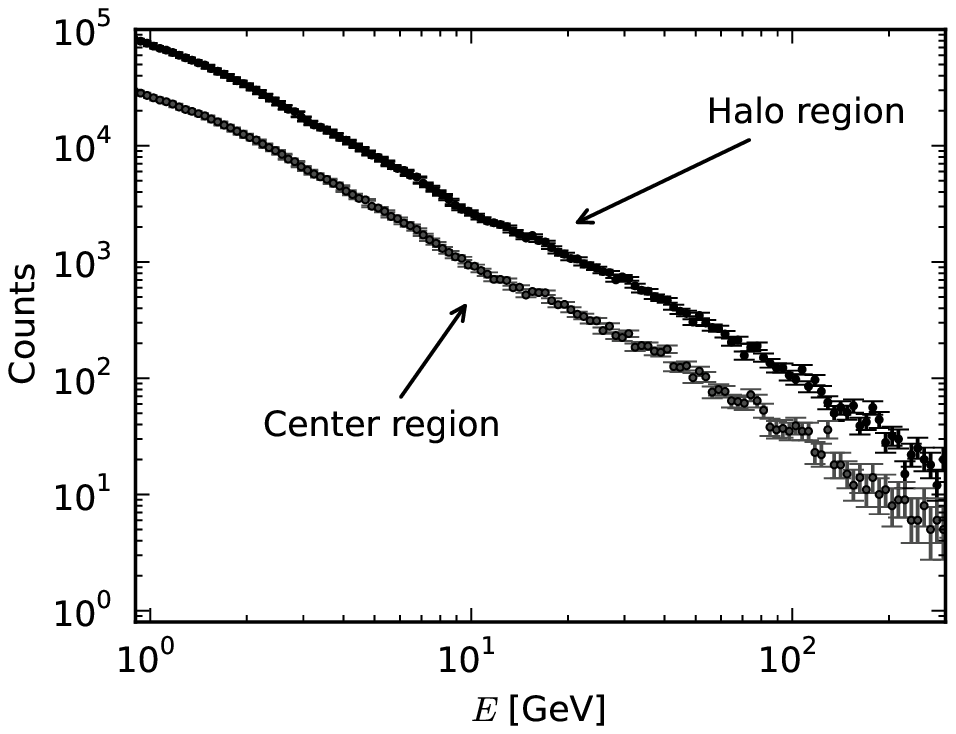}
    \includegraphics[width=0.49\linewidth]{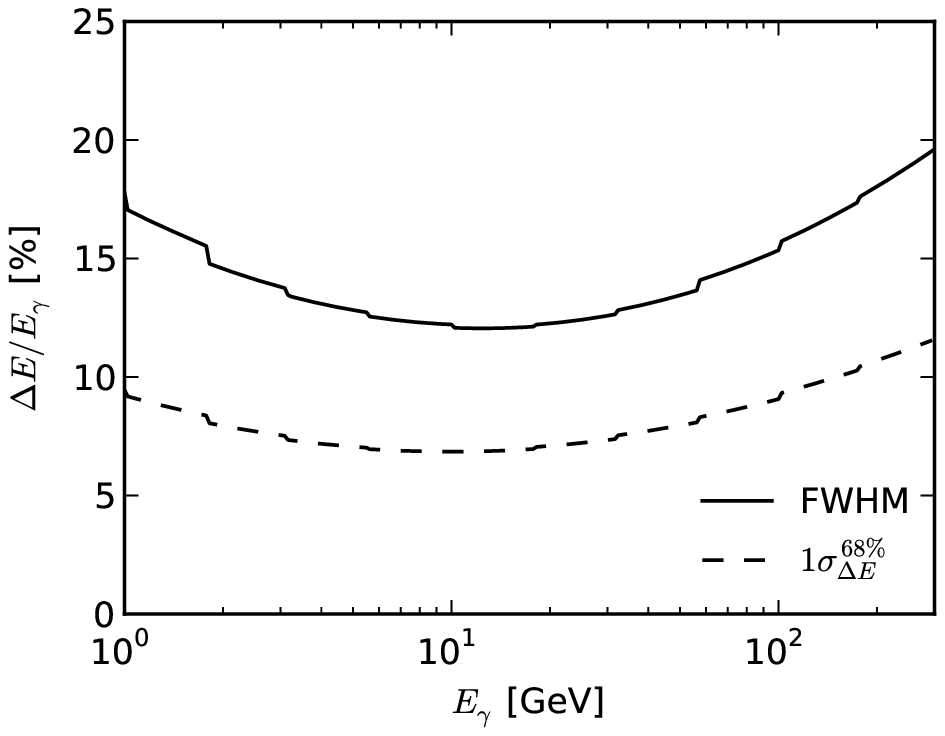}
    \vspace{-0.6cm}
  \end{center}
  \caption{\emph{Left panel:} Observed counts in the halo and the center
  regions, binned into 50 bins per energy decade (200 bins per decade are used
  in our likelihood analysis). \emph{Right panel:} Fermi LAT energy resolution
  $\Delta E/E_\gamma$, in terms of FWHM and 68\% containment $\sigma_{\Delta
  E}^{68\%}$, as derived from the IRF of the `DataClean' event class.}
  \label{fig:counts}
\end{figure}

Since the selection procedure leaves us with a large number of $1.3\times10^6$
($5\times10^5$) events above 1 GeV for the halo (center) region, we perform a
binned analysis of the data. To this end, we distribute the events into 200
logarithmically equally spaced energy bins per decade, and sum over the
angles. This gives a sequence of count numbers $c_i\in\mathbb{N}_0$, which is
illustrated in the left panel of Fig.~\ref{fig:counts} for both target
regions.  Note that we do not perform a point source subtraction in this work.
A proper treatment would mask out only ${\cal O}(5\%)$ of the
events~\cite{Abdo:2010nc}, and hence only marginally affect our results.

The spectral feature produced by a gamma-ray line can be inferred from the
Fermi LAT instrument response function (IRF). Its most recent version, `Pass6
version 3', was determined using Monte Carlo generated samples of photon
events between 18\,MeV and 562\,GeV, and includes effects measured in-flight,
see Refs.~\cite{Rando:2009yq, Abdo:2009gy}. It contains the
point-spread-function (PSF), as well as the energy dispersion $\mathcal{D}(E,
E_\gamma)$ which describes the distribution of the reconstructed energies $E$
as a function of the physical photon energy $E_\gamma$.  In order to integrate
out the implicit dependence of the energy dispersion on the event impact angle
with respect to the detector axis, $\mathcal{D}(E, E_\gamma)$ is averaged over
this impact angle weighted by its distribution in our data sample. The
resulting full width at half maximum (FWHM) of the energy dispersion is shown
in the right panel of Fig.~\ref{fig:counts}, where the $68\%$ containment
energy resolution $\sigma_{\Delta E}^{68\%}$ is also shown.\footnote{Our
resulting FWHM is larger than the one discussed in Ref.~\cite{Abdo:2010nc} by
a factor of $1.1$--$1.4$, which presumably affects our limits at the level of
$5\%$--$20\%$. The difference is probably due to different energy
reconstruction methods underlying the public `DataClean' event class used in
this work and the dedicated analysis in Ref.~\cite{Abdo:2010nc}. However, the
$68\%$ containment agrees well with Ref.~\cite{Atwood:2009ez}.} Lastly, since
our two samples have a large sky coverage, the effects of the PSF, which
features a $68\%$-containment angle well below $1^\circ$ at energies above
1\,GeV, can be safely neglected.\medskip

\begin{figure}[t]
  \begin{center}
    \includegraphics[width=0.49\linewidth]{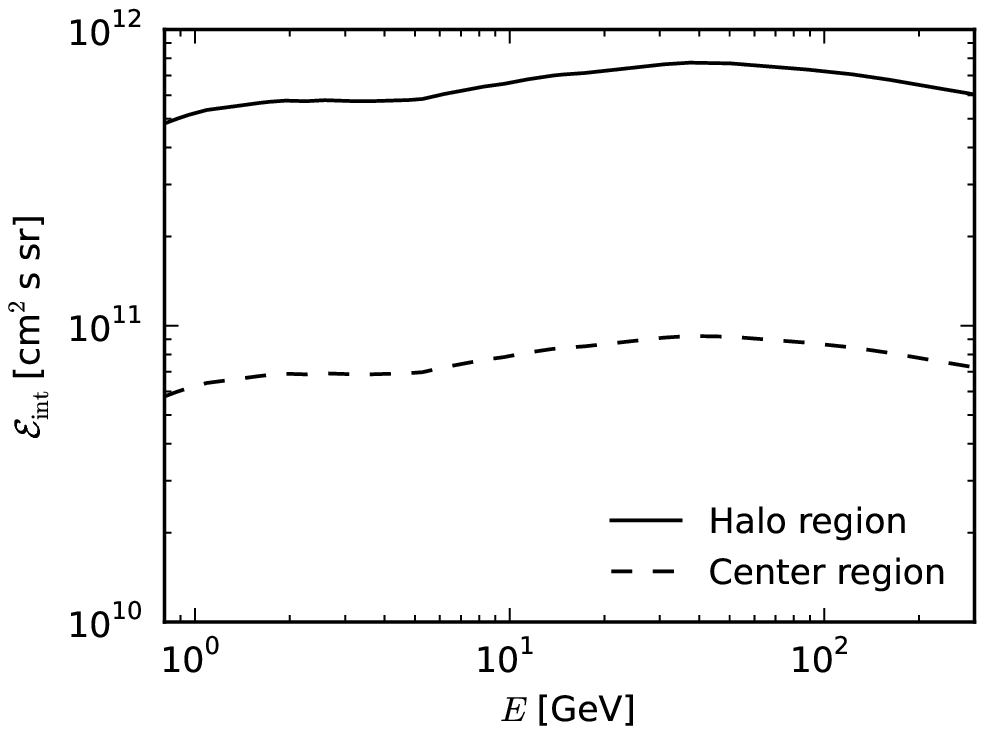}
    \includegraphics[width=0.49\linewidth]{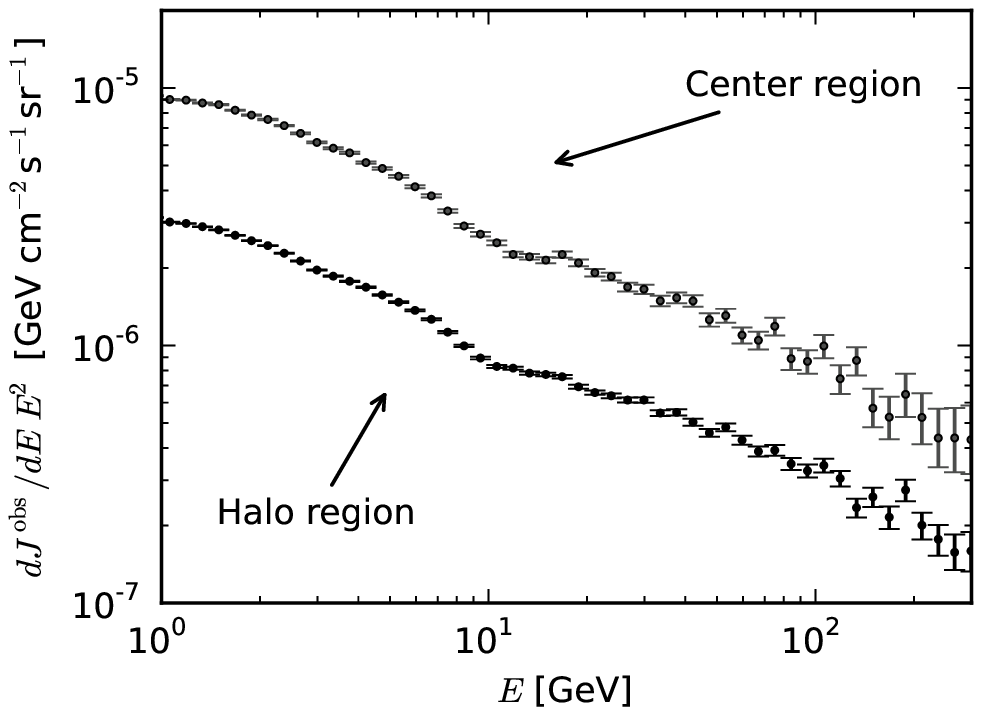}
    \vspace{-0.6cm}
  \end{center}
  \caption{\emph{Left panel:} Integrated exposures $\mathcal{E}_\text{int}$
  for both target regions, following the weighting method described in the
  text.  \emph{Right panel:} Differential gamma-ray flux as observed in the
  two target regions.}
  \label{fig:flux}
\end{figure}

The profile likelihood method~\cite{Rolke:2004mj,Conrad:2007zza} is used to
calculate the significance of a potential gamma-ray line contribution to the
observed gamma-ray flux. The likelihood function is defined as
\begin{align}
  L(\fatmu|{\fatc}) = \Pi_{i=i_0, \dots, i_1} P_{\mu_i}(c_i)\;,
  \label{eq:likelihood}
\end{align}
where, for each energy bin $i$, $\mu_i$ and $c_i$ denote the expected and
observed count numbers, respectively, and $P_\mu(k)=\mu^k e^{-\mu}/k!$ is the
Poisson probability mass function with mean value $\mu$. The data are modeled
by a simple power law with a spectral index $-\gamma$ and a normalization
$\beta$, plus a line signal at fixed energy $E_\gamma$ with the normalization
$\alpha$. The power law background is supposed to take into account not only
the gamma-ray flux of astrophysical origin, but also the gamma-ray continuum
which might be produced in the dark matter annihilation or decay additionally
to the sharp gamma-ray line.  Since the power law is only locally a good
approximation to the background fluxes, we use a sliding energy window in the
fitting procedure. For gamma-ray energies $E_\gamma\lesssim10\GeV$ we only
take into account the energy bins $i\in[i_0, i_1]$ that lie within
$\pm2\sigma_{\Delta E}^{68\%}$ around $E_\gamma$, whereas, due to limited
statistics, at high energies $E_\gamma\gtrsim100\GeV$ bins within a wider
energy window from roughly $\frac13E_\gamma$ to $3 E_\gamma$ are considered.
In the range $10\GeV\lesssim E_\gamma \lesssim 100\GeV$ the energy window size
is interpolated between these two cases.  Furthermore, events above $300\GeV$
are always excluded from the analysis, which leads to energy windows
asymmetric around the highest gamma-ray line energies.
The expected number of counts in each energy bin is given by
\begin{equation}
  \mu_i(\alpha, \beta, \gamma) = \int_{E_i^-}^{E_i^+}dE'\
  \left(\beta\mathcal{E}_\text{int}(E')
  E'^{-\gamma} +\alpha \mathcal{E}_\text{int}(E_\gamma)\mathcal{D}(E',
  E_\gamma) \right)\;,
  \label{eqn:counts}
\end{equation}
where the $E_i^\pm$ denote the boundaries of each energy bin $i$,
$\mathcal{E}_\text{int}$ is the energy dependent integrated total exposure,
and we neglect the effects of the energy dispersion on the power-law
background flux. To obtain $\mathcal{E}_\text{int}$, we integrate the exposure
maps, which we derived with the Fermi Science Tools,\footnote{In
\texttt{gtltcube} we use a zenith cut \texttt{zmax=105} to properly calculate
the exposure also in cases where the field-of-view of the satellite intersects
with the Earth's Albedo. However, due to our above cut on the satellite
rocking angle (see footnote~\ref{fn:gti}) this has only a small effect on the
total exposure.} over the observed region of the sky, weighted by the angular
profile of the expected gamma-ray line signal from dark matter decaying or
annihilating inside the Galactic halo. The resulting integrated exposures are
practically identical for decaying and annihilating dark matter signals, and
are shown in the left panel of Fig.~\ref{fig:flux} for the halo and center
regions. They can be approximated by $\mathcal{E}_\text{int}(E)\simeq
\langle\mathcal{E}(E)\rangle \Delta\Omega$, where the target region averaged
exposures are in the range $\langle\mathcal{E}(E)\rangle\simeq 5$--$8
\times10^{10}\cm^2\s$, and $\Delta\Omega=10.4$\,sr (1.30\,sr) is the solid
angle spanned by the halo (center) region. The observed differential gamma-ray
fluxes for both target regions, as inferred from the exposure and count maps,
are shown in the right panel of Fig.~\ref{fig:flux}.

Depending on whether the normalization $\alpha$ of the gamma-ray line is kept
free or fixed, the likelihood function in Eq.~\eqref{eq:likelihood} must be
maximized with respect to three or two parameters, respectively:
\begin{align}
  L(\fatc)=\sup\{L(\fatmu|\fatc): \alpha, \beta, \gamma\in\mathbb{R}^+\}
  \qquad \text{or} \qquad  L_\alpha(\fatc)=\sup\{L(\fatmu|\fatc):\beta,
  \gamma\in\mathbb{R}^+\}\;.
\end{align}
Defining $\Lambda_\alpha(\fatc) \equiv L_\alpha(\fatc)/L(\fatc)$, the $95\%$
C.L.~interval upper limit of the gamma-ray line normalization is given by the
largest value of $\alpha$ which satisfies $-2\log\Lambda_\alpha(\fatc)\leq4$.
The parameter $\alpha$ is defined such that it can be directly identified with
the corresponding gamma-ray line flux, \textit{i.e.}~$J_\gamma\equiv\alpha$.
In the absence of a gamma-ray line signal at the 5$\sigma$ level, which would
correspond to $-2\log \Lambda_{\alpha=0}(\fatc)\geq23.7$, we quote the upper
limit of the $95\%$ C.L.~interval as the upper limit on the gamma-ray line
flux.  Note that the $-2\log\Lambda_{\alpha=0}$ value corresponding to a
$5\sigma$-confidence level slightly deviates from the naively expected 25,
because the normalization of the gamma-ray line is bounded from below,
\textit{i.e.} $\alpha\geq0$. The adopted value can be derived by recognizing
that in absence of a gamma-ray line in $50\%$ of the cases the optimal fit
would actually require an unphysical negative normalization of the gamma-ray
line, since for large enough count numbers statistical fluctuations are
symmetric around their mean.  These cases yield $-2\log\Lambda_{\alpha=0}=0$,
whereas in the other $50\%$ of the cases, the $-2\log\Lambda_{\alpha=0}$ would
follow a conventional $\chi^2$-distribution with one degree of
freedom.\footnote{We confirmed this behavior with a Monte Carlo analysis.
Details about the likelihood-ratio and confidence intervals can be found \fex
in Refs.~\cite{Wilks:1938, Feldman:1997qc}. Note that we do not use a ``trial
factor'' as discussed in Refs.~\cite{Edmonds:2007zz, Ylinen:2010zz}, which
only becomes relevant when constraining models that predict a multitude of
lines or when claiming a line discovery.}

The likelihood analysis is performed using PyFITS\footnote{PyFITS is a product
of the Space Telescope Science Institute, which is operated by AURA for NASA,
see \url{http://www.stsci.edu/resources/software_hardware/pyfits}.} and
PyMinuit, an interface to the function minimizer MINUIT~1.7.9.\footnote{See
\url{http://code.google.com/p/pyminuit/} and
\url{http://seal.web.cern.ch/seal/snapshot/work-packages/mathlibs/minuit/}.}

\begin{figure}[t]
  \begin{center}
    \includegraphics[width=0.9\linewidth]{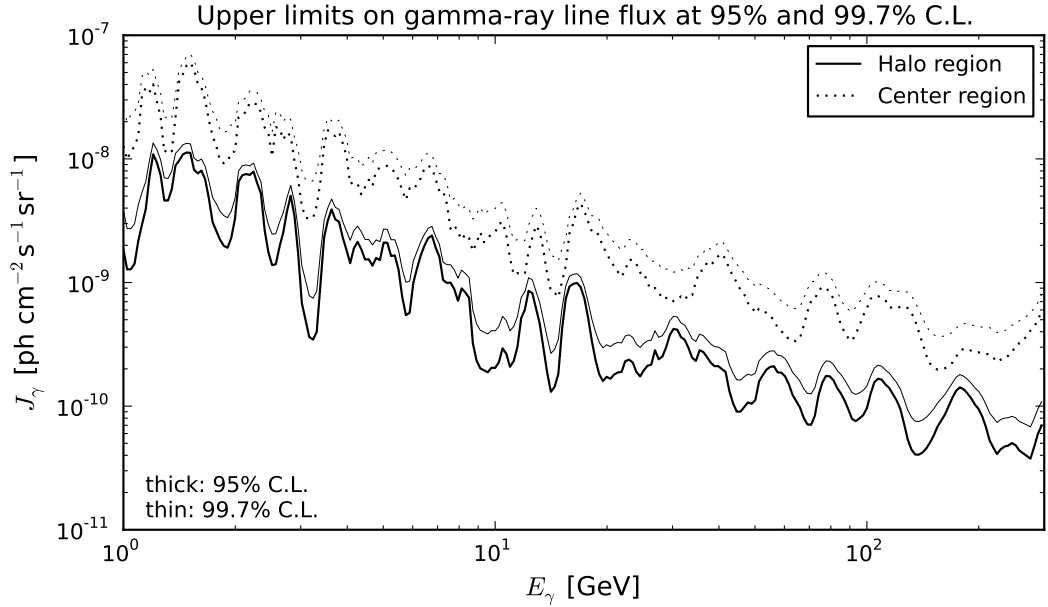}
    \vspace{-0.4cm}
  \end{center}
  \caption{Our results for the $95\%$ and $99.7\%$ C.L. upper limits on the
  monochromatic photon flux from the halo and the center regions, as function
  of the gamma-ray line energy $E_\gamma$. Note that due to instrumental
  effects these limits are expected to be too strong at lower energies in some
  cases (cf. section~\ref{sec:discussion}).  The corresponding conservative
  limits are show in Tab.~\ref{tab:bounds} (as well as in the
  Figs.~\ref{fig:bounds_DDM} and \ref{fig:bounds_ADM}).}
  \label{fig:bounds_flux}
\end{figure}

\subsection{Results}
\label{sec:results}
No gamma-ray lines with $5\sigma$ significance are found in the 1--300\,GeV
energy range.  Details and caveats of the analysis are discussed in
section~\ref{sec:discussion}.  Our results for the $95\%$ and $99.7\%$
C.L.~limits on the gamma-ray line flux are shown in Fig.~\ref{fig:bounds_flux}
as a function of the gamma-ray line energy for both halo and center regions.
As discussed in section~\ref{sec:discussion}, instrumental effects forbid us
to take these limits at face value; the conservative limits summarized in
Tab.~\ref{tab:bounds} should be used instead.  These are derived by quoting
the weakest limit we obtain in given energy bands. Together with
Eqs.~\eqref{eqn:fluxDDM} and \eqref{eqn:fluxADM}, it is then straightforward
to derive the corresponding constraints on the dark matter decay width or
annihilation cross section.

\begin{table}[t]
  \centering
  \begin{tabular}{@{}ccccccccc@{}}
    \toprule
    Target region &\phantom{a}&
    \multicolumn{3}{c}{
    $\Gamma_{\psi\to\gamma\nu}^{\text{profile}}/
    \Gamma_{\psi\to\gamma\nu}$
    } &
    \phantom{a} &
    \multicolumn{3}{c}{
    $\langle\sigma
    v\rangle_{\psi\psi\to\gamma\gamma}^{\text{profile}}/\langle\sigma
    v\rangle_{\psi\psi\to\gamma\gamma}$
    }\\
    \cmidrule{3-5} \cmidrule{7-9}
    && Iso. & NFW & Ein. && Iso. & NFW & Ein.\\
    \midrule
    Halo   && 0.96 & 1.00 & 1.00 && 1.03 & 1.00 & 0.89 \\
    Center && 1.03 & 1.00 & 0.91 && 1.65 & 1.00 & 0.64 \\
    \bottomrule
  \end{tabular}
  \caption{Rescaling factors for the bounds on the dark matter decay width and
  annihilation cross section (shown in Tab.~\ref{tab:bounds}), corresponding
  to the different halo profiles. In most cases, the weakest bounds are
  obtained for the isothermal profile (Iso.), the strongest bounds for the
  Einasto profile (Ein.), whereas the NFW profile leads to intermediate
  constraints.}
  \label{tab:rescaling}
\end{table}

\begin{figure}[t]
  \begin{center}
    \includegraphics[width=0.9\linewidth]{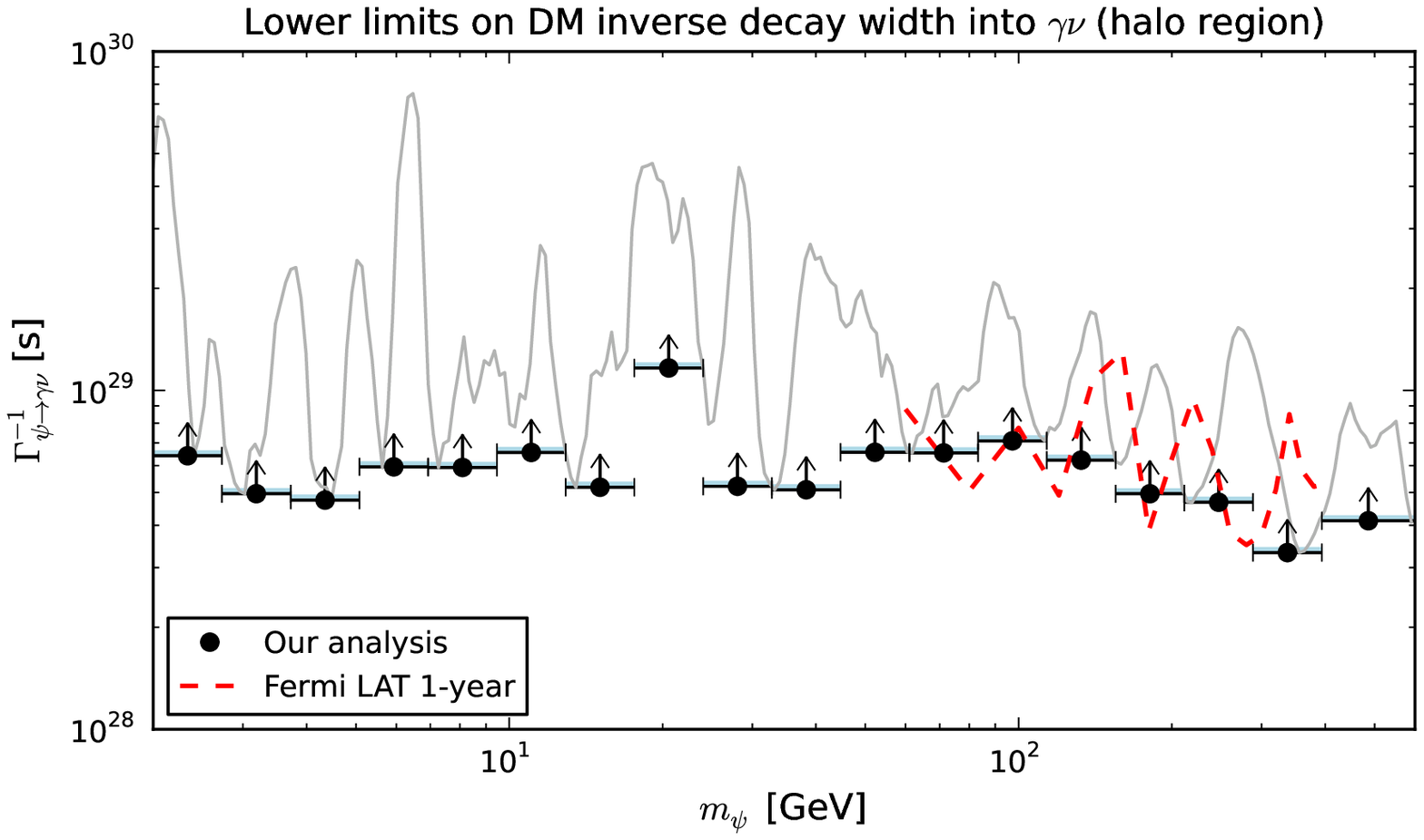}
    \vspace{-0.4cm}
  \end{center}
  \caption{Lower bounds on the dark matter inverse decay width into
  monochromatic photons and neutrinos, $\Gamma_{\psi\to\gamma\nu}^{-1}$, as a
  function of the dark matter mass $m_\psi$, derived from the halo region
  fluxes assuming the NFW dark matter profile. The gray-solid line shows the
  $95\%$ C.L.~limits as directly derived from the line flux limits shown in
  Fig.~\ref{fig:bounds_flux}. The black dots show the weakest limits obtained
  in the adopted energy bands and are listed in Tab.~\ref{tab:bounds}. For
  comparison, the previous Fermi LAT limits from Ref.~\cite{Abdo:2010nc} are
  also shown by the red-dashed line (rescaled to our decay channel). The
  barely visible blue bands illustrate how the bounds change when using the
  isothermal or Einasto dark matter profiles instead.}
  \label{fig:bounds_DDM}
\end{figure}

Limits on the inverse dark matter decay width into monochromatic photons and
neutrinos, $\Gamma_{\psi\to\gamma\nu}^{-1}$, as derived from the halo region
observations, are presented in Fig.~\ref{fig:bounds_DDM}, assuming a NFW dark
matter profile. The gray line shows the bounds as they directly follow from
the flux limits in Fig.~\ref{fig:bounds_flux}. In most of the considered
energy range, they fluctuate around $10^{29}$\,s. The slight improvement with
respect to the previous Fermi LAT limits~\cite{Abdo:2010nc} (dashed line) is
due to the increased statistics after two years of data taking.  The black
dots represent the weakest limits obtained when varying the gamma-ray line
energy within different adopted energy bands. Except at the highest energies,
these values fluctuate around $6\times10^{28}$\,s, and are summarized in
Tab.~\ref{tab:bounds}. Adopting halo profiles different from the NFW profile
would change the limits by a few percent. The corresponding rescaling factors
for the Einasto and the isothermal profiles are summarized in
Tab.~\ref{tab:rescaling}, and their impact on the limits is illustrated by the
blue band in Fig.~\ref{fig:bounds_DDM}. Note that the normalization of the
dark matter profile at Sun's position is another source of uncertainty, which
is not captured in the rescaling parameters of Tab.~\ref{tab:rescaling}.  We
do not plot the somewhat weaker limits on the decay width obtained from the
center region, which can be found in Tab.~\ref{tab:bounds} for completeness.

\begin{figure}[t]
  \begin{center}
    \includegraphics[width=0.9\linewidth]{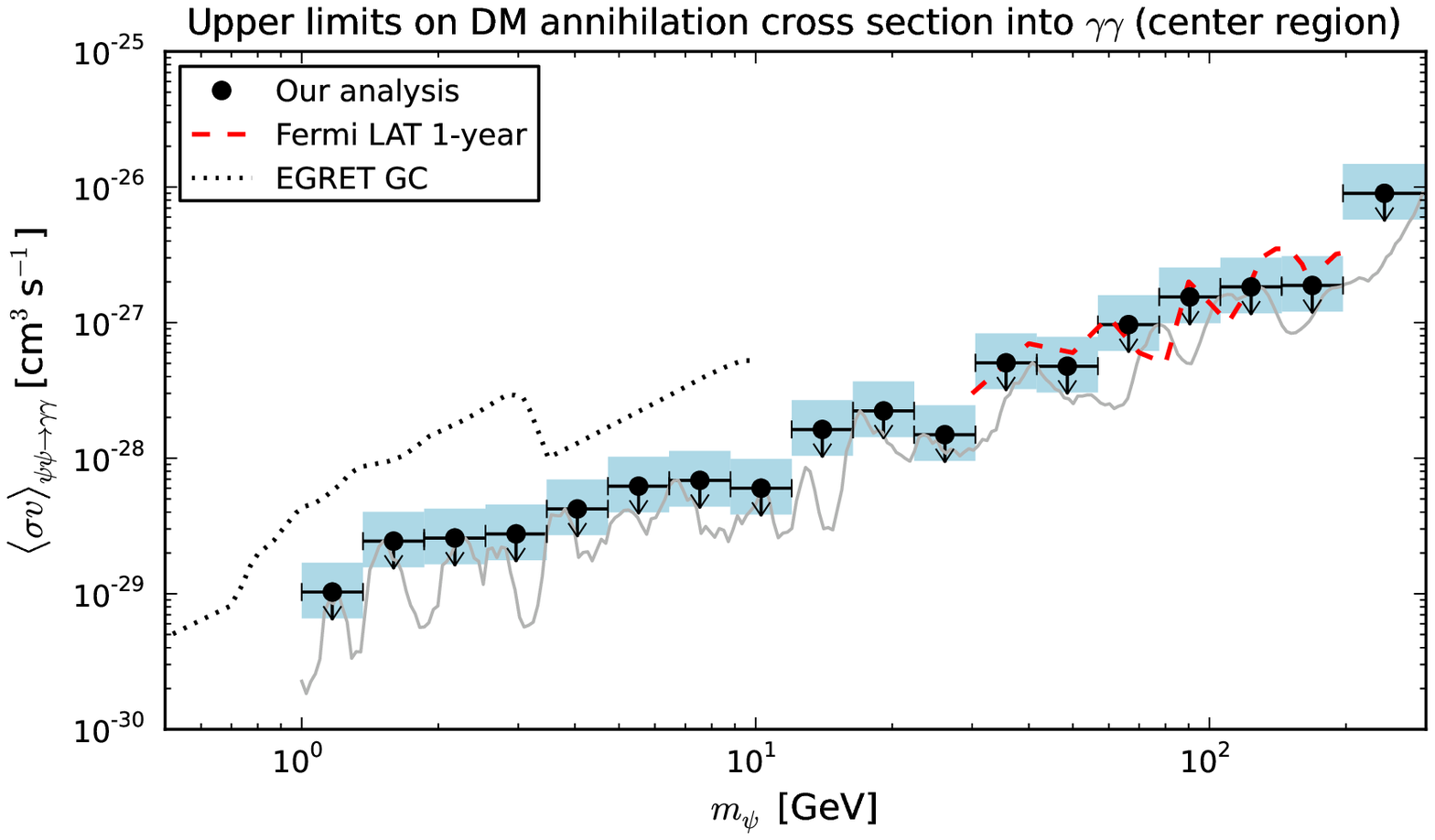}
    \vspace{-0.4cm}
  \end{center}
  \caption{Upper bounds on the annihilation cross section into gamma-pairs,
  $\langle\sigma v\rangle_{\psi\psi\to\gamma\gamma}$, as a function of the
  dark matter mass $m_\psi$, derived from the center region fluxes assuming
  the NFW dark matter profile. The gray-solid line shows the $95\%$
  C.L.~limits as directly derived from the line flux limits shown in
  Fig.~\ref{fig:bounds_flux}. The black dots show the weakest limits obtained
  in the adopted energy bands and are listed in Tab.~\ref{tab:bounds}. For
  comparison, the previous Fermi LAT limits from Ref.~\cite{Abdo:2010nc} as
  well as the limits derived from EGRET observations of the Galactic
  center~\cite{Pullen:2006sy} are also shown by the red-dashed and the
  black-dotted lines, respectively. The blue bands illustrate how the bounds
  change when using the isothermal or Einasto dark matter profiles instead.}
  \label{fig:bounds_ADM}
\end{figure}

In Fig.~\ref{fig:bounds_ADM} we present the limits on the dark matter
annihilation cross section into gamma-ray pairs, $\langle \sigma
v\rangle_{\psi\psi\to\gamma\gamma}$, as derived from the center region
observations. As illustrated by the blue band, the uncertainties coming from
the dark matter density profile are much larger than in the case of dark
matter decay and of the order of $\sim50\%$,
\textit{cf.}~Tab.~\ref{tab:rescaling}. For comparison, we also show the
previous limits obtained by EGRET~\cite{Pullen:2006sy}.\footnote{Note that the
cross section limits shown in Fig.~8 of Ref.~\cite{Pullen:2006sy} are
incorrectly too strong by a factor of ten, whereas the flux limits presented
in Fig.~7 are correct~\citetext{priv.~comm.~with the authors}. The EGRET cross
section limits that we present in Fig.~\ref{fig:bounds_ADM} are derived from
Fig.~7 of Ref.~\cite{Pullen:2006sy}.} The present analysis improves these
limits by up to an order of magnitude, which is essentially due to the
increased statistics of Fermi LAT with respect to EGRET. Furthermore, at high
energies, we slightly strengthen the previous Fermi LAT
constraints~\cite{Abdo:2010nc}. Our results are summarized in
Tab.~\ref{tab:bounds}. It is interesting to note that, assuming the NFW
profile, the limits on the dark matter annihilation cross section derived from
the halo region are only a factor $\sim2$ weaker than the limits derived from
the center region.\medskip

Using kinematic considerations, our results can be straightforwardly
translated into constraints on two-body decay or annihilation into a
monochromatic photon plus a massive neutral particle $N$ with mass $m_N$. This
is relevant for the common case of annihilation/decay into $\gamma Z^0$.
However, in principle $m_N$ is a free parameter, and if $m_N$ is close to the
dark matter mass $m_\psi$, the line signal can be even strongly enhanced in
some scenarios~\cite{Garny:2010eg}. Defining the quantity
\begin{align}
  x\equiv1+\sqrt{1+\frac{m_N^2}{E_\gamma^2}}\;,
\end{align}
which is bounded from below by $x\geq2$, our limits for the decay into
$\gamma\nu$ translate into limits on the decay into $\gamma N$ via
\begin{align}
  \Gamma_{\psi\to\gamma N}(m_\psi)=\frac
  x2\,\Gamma_{\psi\to\gamma\nu}(E_\gamma)\;,
  \quad\text{where}\quad
  m_\psi=x\, E_\gamma\;.
\end{align}
For the annihilation scenario the cross section limits on $\psi\psi\to\gamma
N$ are obtained by
\begin{align}
  \langle\sigma v\rangle_{\psi\psi\to\gamma N}(m_\psi)= \frac{x^2}2
  \,\langle\sigma v\rangle_{\psi\psi\to\gamma\gamma}(E_\gamma)\;,
  \quad\text{where}\quad
  m_\psi = \frac x2\, E_\gamma\;.
\end{align}
\medskip

\medskip
Models that are partially constrained by our limits include singlet dark
matter~\cite{Profumo:2010kp} and SU(2) vector dark matter~\cite{Arina:2009uq},
but also Refs.~\cite{Mambrini:2009ad, Mambrini:2010yp, Goodman:2010qn,
Jackson:2009kg,Garny:2010eg}. Scenarios that remain practically unconstrained
include the standard MSSM scenario~\cite{Bergstrom:1997fj}, but also
Refs.~\cite{Profumo:2008yg, Ferrer:2006hy, Gustafsson:2007pc, Bertone:2009cb,
Bertone:2010fn}. The particular case of gravitino dark matter will be
discussed in the next section.

\subsection{Discussion}
\label{sec:discussion}
\begin{figure}[t]
  \begin{center}
    \includegraphics[width=0.49\linewidth]{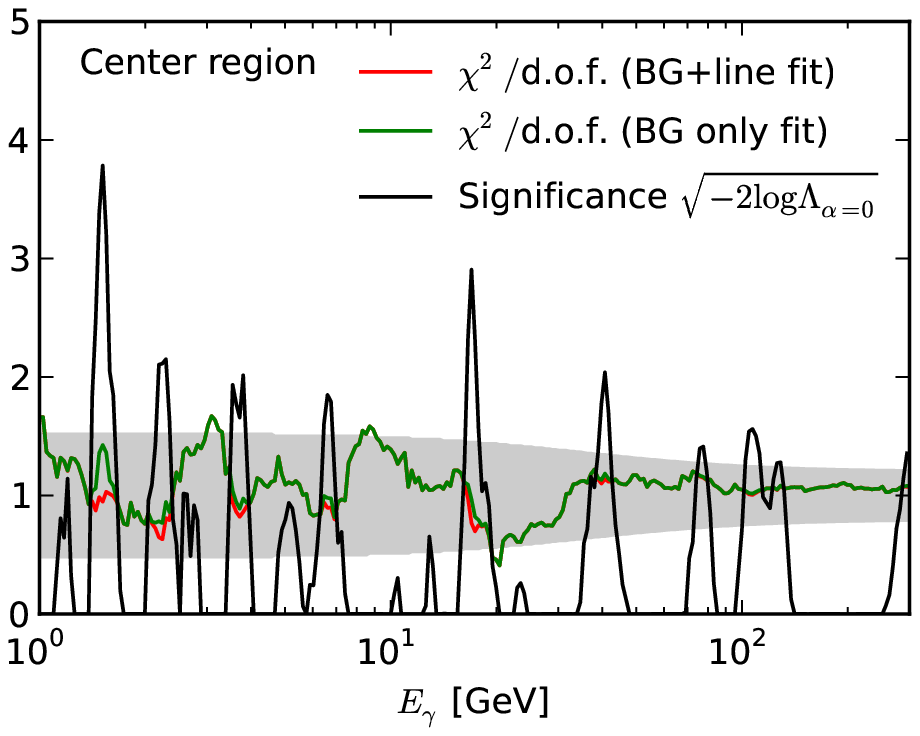}
    \includegraphics[width=0.49\linewidth]{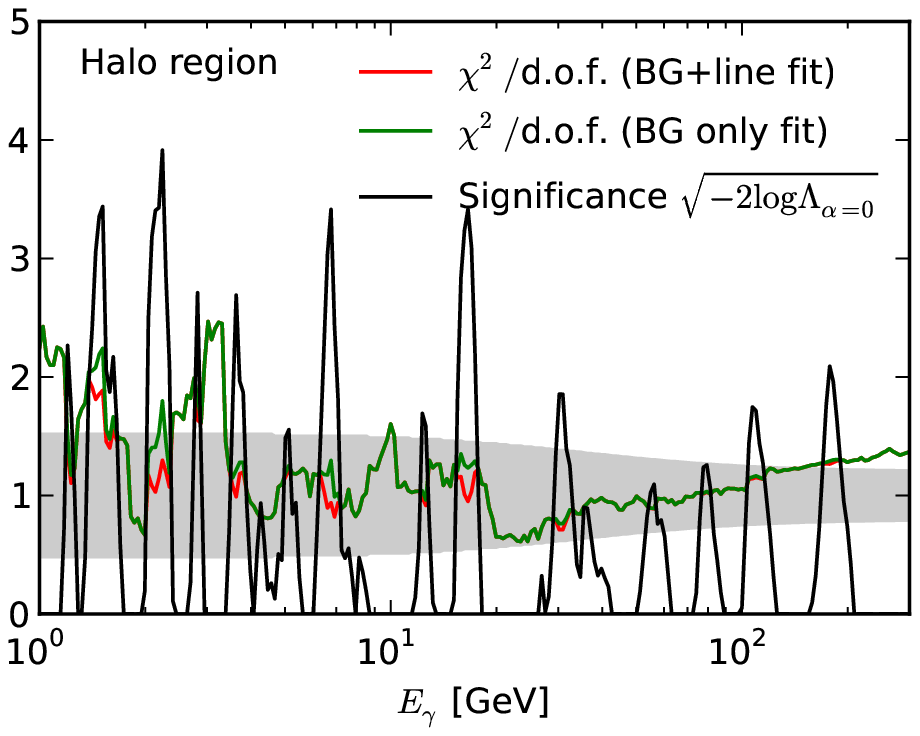}
    \vspace{-0.6cm}
  \end{center}
  \caption{Significance of a potential line signal (black line) in the center
  (left panel) and halo (right panel) region in terms of standard deviations.
  We also show the $\chi^2$/d.o.f.~values of fits with (red line) and without
  (green line) line signal. The gray shaded area corresponds to a p-values of
  $\geq5\%$. Especially at low energies in the halo region, significant
  deviations occur from what is statistically expected in absence of a signal,
  as discussed in section~\ref{sec:discussion}.}
  \label{fig:lambda}
\end{figure}

One crucial assumption underlying our analysis is that the background flux in
the different considered energy windows can be well approximated by a
power-law. This assumption is most likely to break down in cases where the
statistics is very good. In order to check the validity of a power-law ansatz,
we show in Fig.~\ref{fig:lambda} the $\chi^2$/d.o.f.~of the background-only
(green lines) and of the background-plus-signal (red lines) fits, as function
of the gamma-ray line energy.\footnote{The smallness of the differences
between the $\chi^2$/d.o.f.~of the background-plus-signal and background-only
fit at high energies comes from the fact that the $\chi^2$ values are actually
dominated by the background and not by the narrow signal.} The grey band
corresponds to a $p$-value of $\geq5\%$.  For the center region the fits are
essentially in agreement with the data over the whole energy range.  However,
$p$-values significantly smaller than $5\%$ occur at energies between $1$ and
$10\GeV$ (as well as at high energies close to $300\GeV$) when considering the
halo region, which has a three times larger statistics than the center region.
Assuming that the astrophysical gamma-ray fluxes follow smooth bended
power-laws, this tension points to an instrumental effect, presumably related
to the energy reconstruction of gamma-ray events.

The black lines in Fig.~\ref{fig:lambda} show the significance (in terms of
standard deviations) of line signals with respect to a power-law background.
Over the whole energy range we are performing $\mathcal{O}(50)$ statistically
independent tests for a line.  Hence, we expect to see a few cases with
$2\sigma$ significance, and practically no case with $3\sigma$ significance.
This expectation is marginally compatible with the observations of the center
region, and is clearly violated for the halo region, where four $>3\sigma$
excesses are observed below $20\GeV$. However, an interpretation of these
excesses in terms of gamma-ray lines would be suspicious: They occur at low
energies in the halo region, where the statistics is best and hence the impact
of instrumental effects most relevant, as indeed indicated by the bad
$\chi^2$/d.o.f.~of the background-fits at this energy range. Moreover, most
line-like features appear, although with a different significance, at the same
position in both, the halo and the center region, indicating again a
systematic effect. Without further understanding and reduction of instrumental
effects, no conclusive statement about the existence of gamma-ray lines (below
$\sim5\sigma$ significance) in this energy range can be made.

The limits presented in Figs.~\ref{fig:bounds_DDM}, \ref{fig:bounds_ADM} and
Tab.~\ref{tab:bounds} are conservative in the sense that we always quote the
weakest limit we obtain in given energy bands.  Especially in the halo region
at energies below $20\GeV$, this ensures that our limits are dominated (and
weakened) by the presence of the line-like artefacts in the energy spectrum. A
future removal of these artefacts from the data is expected to lead in general
to a strengthening of the presented limits.  Note that the presence of a real
gamma-ray line signal significantly stronger than the artefacts would have
lead to limits dominated by the line signal itself.

\begin{figure}[t]
  \begin{center}
    \includegraphics[width=0.49\linewidth]{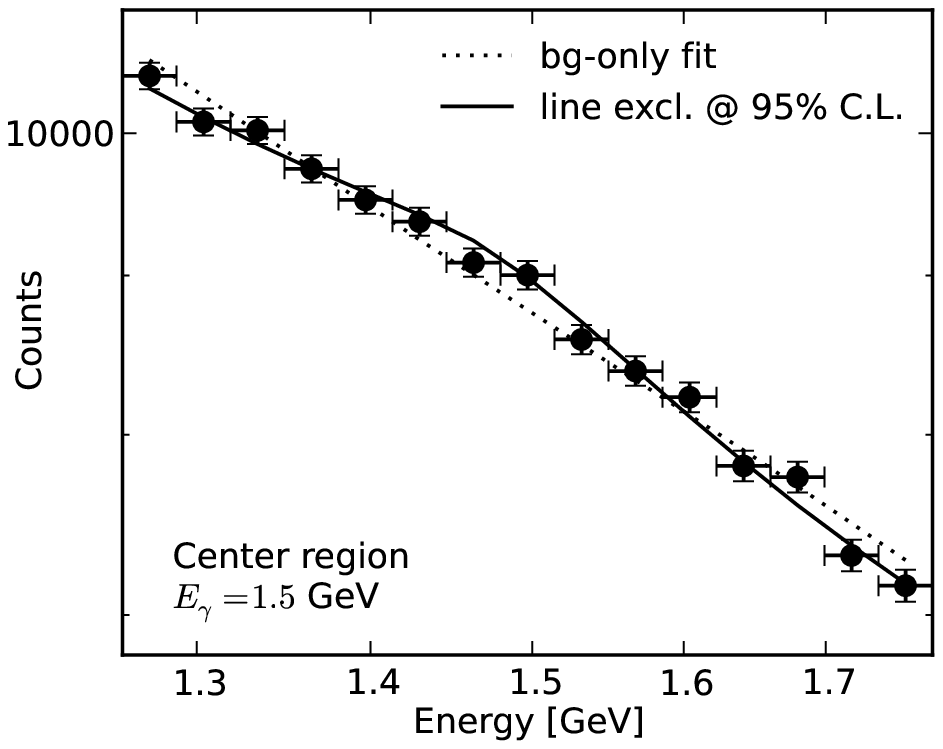}
    \includegraphics[width=0.49\linewidth]{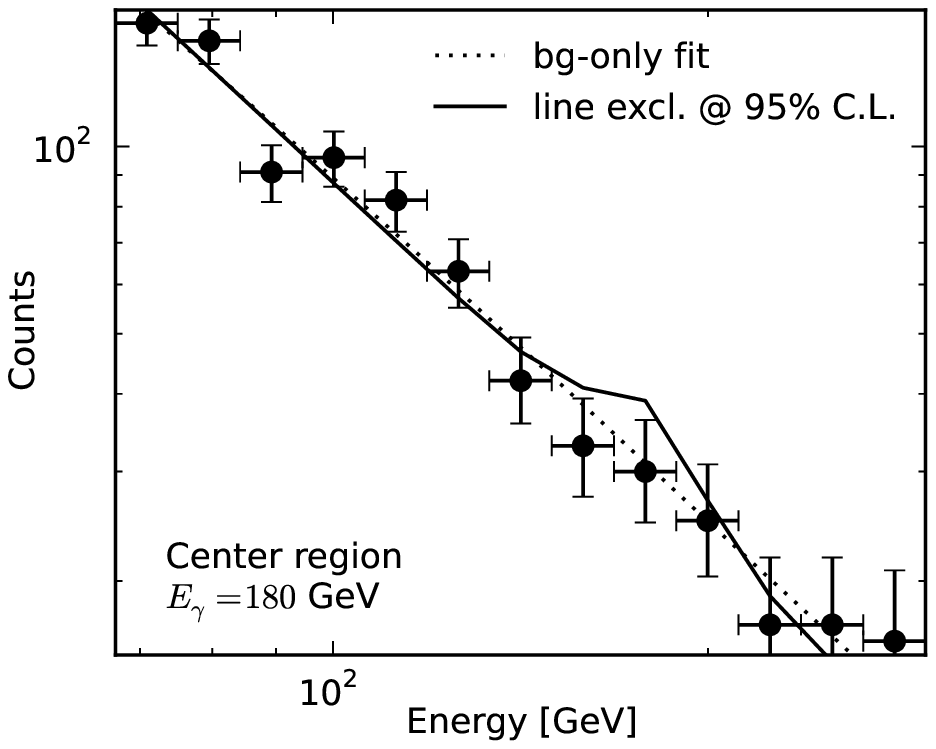}
    \includegraphics[width=0.49\linewidth]{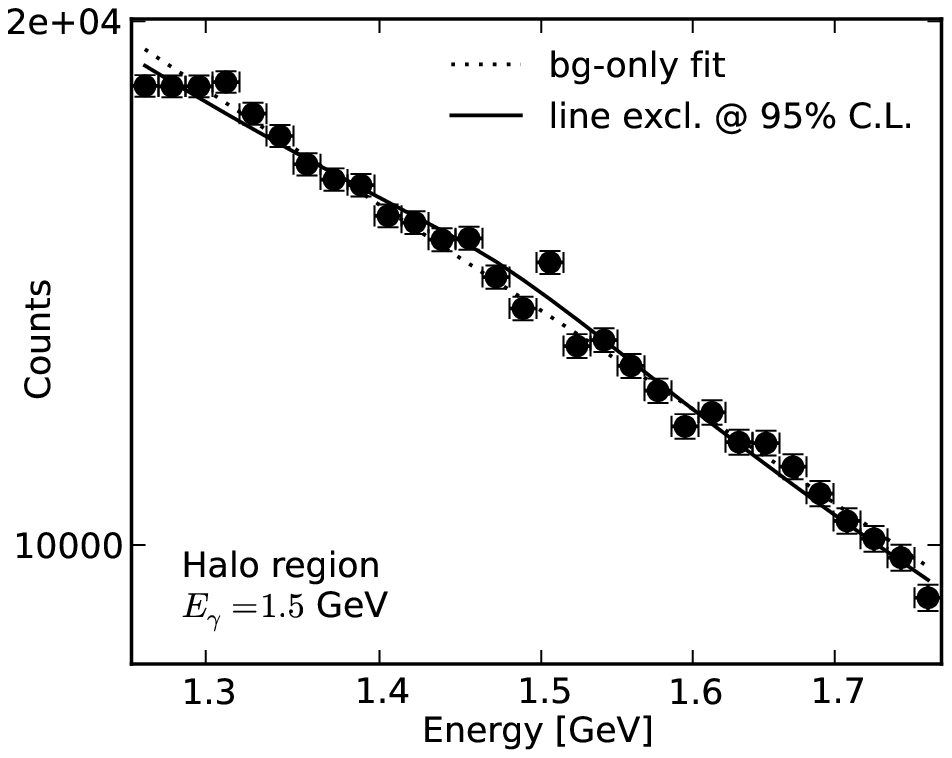}
    \includegraphics[width=0.49\linewidth]{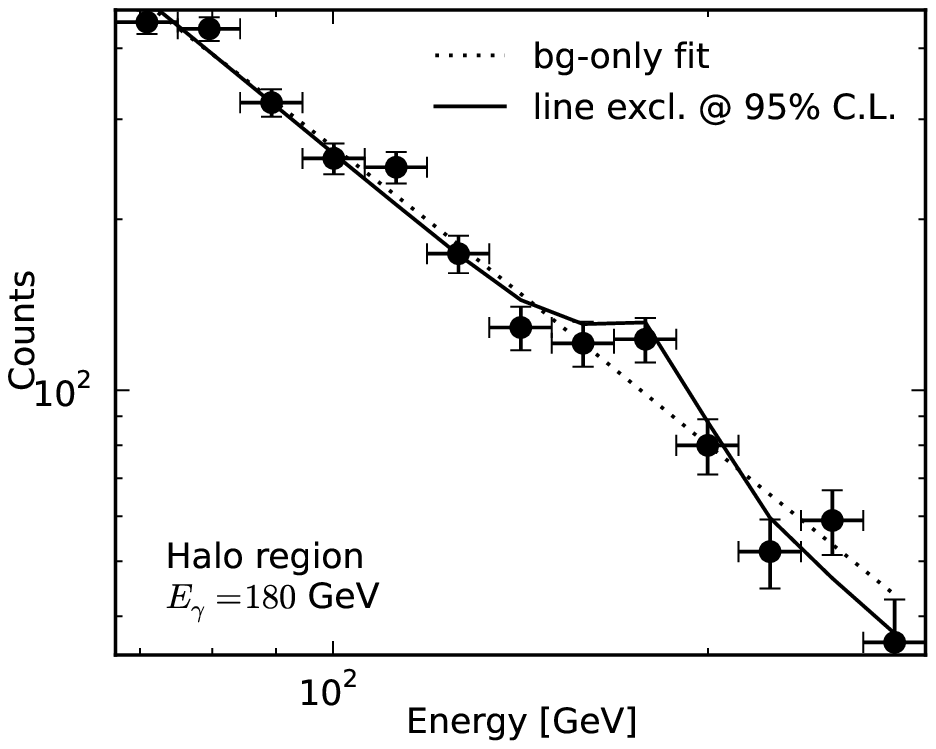}
    \vspace{-0.6cm}
  \end{center}
  \caption{Line signals that are excluded at 95\% C.L. (solid line), compared
  to best-fit background-only model (dotted line) and data. We show results
  for the center (top panels) and halo (bottom panels) region. The gamma-ray
  line energies are $E_\gamma=1.5\GeV$ (left panels) and $E_\gamma=180\GeV$
  (right panels). In the left panels, the signal-to-background ratio is
  $2.6\%$ (top) and $1.5\%$ (bottom).}
  \label{fig:lines}
\end{figure}

\begin{table}[t]
  \centering
  \begin{tabular}{@{}cccccccccc@{}}
    \toprule
    $E_\gamma$ [GeV] &\phantom{a}&
    \multicolumn{2}{c}{$J_\gamma$ [ph/(cm$^{2}$ s sr)]} &
    \phantom{a} &
    \multicolumn{2}{c}{$\Gamma_{\psi\to\gamma\nu}^{-1}$ [s]} &
    \phantom{a} &
    \multicolumn{2}{c}{$\langle\sigma v\rangle_{\psi\psi\to\gamma\gamma}$
    [cm$^3$s$^{-1}$]}\\
    \cmidrule{3-4} \cmidrule{6-7} \cmidrule{9-10}
    && halo & center && halo & center && halo & center\\
    && $\times10^{-9}$ & $\times10^{-9}$ && $\times10^{28}$&$\times10^{28}$
    && $\times10^{-28}$ &$\times10^{-28}$ \\
    \midrule
    1.0--1.4 && 11 & 40 && 6.4 & 4.4 && 0.25 & 0.10 \\
    1.4--1.9 && 11 & 60 && 5.0 & 2.3 && 0.41 & 0.24 \\
    1.9--2.5 && 7.9 & 29 && 4.7 & 3.3 && 0.63 & 0.26 \\
    2.5--3.5 && 5.0 & 18 && 5.9 & 4.6 && 0.63 & 0.22 \\
    3.5--4.7 && 3.9 & 17 && 5.9 & 3.4 && 0.82 & 0.42 \\
    4.7--6.5 && 2.1 & 8.8 && 6.6 & 4.7 && 1.3 & 0.49 \\
    6.5--8.8 && 2.4 & 8.8 && 5.2 & 3.6 && 1.7 & 0.69 \\
    8.8--12 && 0.60 & 3.1 && 12 & 6.6 && 1.4 & 0.60 \\
    12--16 && 0.98 & 3.0 && 5.2 & 4.4 && 4.2 & 1.4 \\
    16--22 && 0.99 & 4.4 && 5.1 & 2.9 && 4.4 & 2.2 \\
    22--31 && 0.42 & 1.5 && 6.6 & 5.9 && 6.2 & 1.5 \\
    31--42 && 0.42 & 1.7 && 6.5 & 3.0 && 6.3 & 5.1 \\
    42--57 && 0.21 & 1.5 && 7.1 & 3.3 && 11 & 4.8 \\
    57--77 && 0.20 & 0.92 && 6.2 & 3.0 && 17 & 9.4 \\
    77--106 && 0.18 & 0.90 && 5.0 & 2.6 && 29 & 15 \\
    106--144 && 0.17 & 0.78 && 4.7 & 2.6 && 31 & 18 \\
    144--197 && 0.14 & 0.29 && 3.3 & 3.9 && 72 & 18 \\
    197--300 && 0.10 & 0.59 && 4.1 & 1.2 && 96 & 91 \\
    \bottomrule
  \end{tabular}
  \caption{Conservative upper limits on the flux in gamma-ray lines
  $J_\gamma$, as well as on the decay width $\Gamma_{\psi\to\gamma\nu}$ and
  the annihilation cross section $\langle\sigma
  v\rangle_{\psi\psi\to\gamma\gamma}$ of dark matter into monochromatic
  photons, for line energies $E_\gamma=1$--$300$\,GeV. In each of the
  indicated energy bands, we varied $E_\gamma$ to find the weakest possible
  $95\%$ C.L.~limit. The flux limits for all values of $E_\gamma$ are shown in
  Fig.~\ref{fig:bounds_flux} (see also section 2.3). The limits are evaluated
  for the halo and for the center region separately, and hold for the NFW
  profile. Other dark matter profiles just require a rescaling of the bounds
  as shown in Tab.~\ref{tab:rescaling}.}
  \label{tab:bounds}
\end{table}

For two different gamma-ray line energies and for both center and halo
regions, we show in Fig.~\ref{fig:lines} the line signals that are excluded at
$95\%$ C.L.~(for which $-2\log\Lambda_\alpha(\fatc)=4$, see above), and
compare them with the best-fit background-only fluxes.  The gamma-ray line
energies are chosen such that the corresponding limits dominate in their
respective energy bands in case of the halo region. Furthermore, the energy
ranges shown in the plots correspond to the energy windows used in the fitting
procedure, and we rebinned the data for better visibility.  At low energies
the gamma-ray line constraints essentially come from the non-observation of a
strong bending in the measured energy spectrum (with signal-to-background
ratios as small as $1.5\%$), whereas at high energies, where the considered
energy window is larger, the constraints come from the non-observation of a
well defined bump.  

In summary, due to the apparent presence of line-like artefacts at the
few percent level in the energy spectrum of the public `DataClean' event class
of the Fermi LAT, the flux limits from Fig.~\ref{fig:bounds_flux} are expected
to be too strong in some cases, especially at energies below $\sim20\GeV$.
Instead, the limits presented in Fig.~\ref{fig:bounds_DDM},
\ref{fig:bounds_ADM} and Tab.~\ref{tab:bounds} should be used.

\medskip
Further more general uncertainties and systematic effects enter our
analysis~\cite{Edmonds:2007zz, Fermi:caveats}: The systematic uncertainties in
the effective area are energy dependent and increase from $5\%$ at 560 MeV to
$20\%$ at 10--100 GeV in case of the `DataClean' event
class~\cite{Abdo:2010nz} (note that these uncertainties amount mainly to an
overall shift in the effective area). Uncertainties on the level of around
$20\%$ are expected to continue up to energies of hundreds of
GeV~\cite{Edmonds:2007zz, Fermi:caveats}, but one has to keep in mind that a
precise determination of the effective area at these high energies is still
lacking due to limited statistics. However, even values reasonably larger than
$20\%$ would only mildly affect our conclusions. Over the whole considered
energy range, the systematics of the LAT effective area consequently translate
into uncertainties for our dark matter limits that range between $5\%$ and
$\sim20\%$.  The absolute energy scale of the Fermi LAT is determined up to
$^{+5\%}_{-10\%}$~\cite{Abdo:2009zk}.  This translates into an additional
uncertainty on the dark matter annihilation cross section $\langle \sigma
v\rangle_{\psi\psi\to\gamma\gamma}$ of $^{+10\%}_{-20\%}$, and on the decay
width $\Gamma_{\psi\to\gamma\nu}$ of $^{+5\%}_{-10\%}$, respectively. 

\medskip
Finally, we have also checked that when extending our analysis to $500\GeV$ no
significant line signals at high energies are revealed, and the limits below
the highest energy band presented in Tab.~\ref{tab:bounds} are only marginally
affected. Considering the $197$--$300\GeV$ ($300$--$500\GeV$) energy band, we
obtain in this case $\Gamma^{-1}_{\psi\to\gamma\nu} > 6.0\times10^{28}\s\
(1.7\times10^{28}\s)$ for the halo region, and $\langle\sigma
v\rangle_{\psi\psi\to\gamma\gamma} < 4.1\times10^{-27}\cm^3\s^{-1}\
(1.54\times10^{-26}\cm^3\s^{-1})$ for the center region.  However, we
emphasize that energies above 300\,GeV lie outside of the nominal energy range
of the Fermi LAT.

\section{Consequences for Gravitino Dark Matter}
Local extensions of the supersymmetric standard model predict the existence of
the gravitino, the supersymmetric partner of the graviton.  Depending on the
mechanism of supersymmetry breaking, it can be the lightest superparticle
(LSP). In this case, it provides a natural dark matter
candidate~\cite{Pagels:1981ke}, and is an interesting alternative to the
standard WIMP scenario.

The dominant contribution to the gravitino production in the early Universe
comes from the 2-to-2 QCD scatterings, which are taking place in the thermal
bath at reheating. The resulting energy density is given by 
\begin{align}
  \Omega_{3/2}^{\text{th}}\, h^2 = C\,\left( \frac{100 \,\text{GeV}}{m_{3/2}}
  \right)
  \bigg( \frac{m_{\tilde g}}{1 \,\text{TeV}}\bigg)^2 \left(
  \frac{T_R}{10^{10}\,\text{GeV}} \right)\,,
  \label{eq:gravitinoabundance}
\end{align}
where $T_R$ is the reheating temperature, while $m_{3/2}$ and $m_{\tilde g}$
are the gravitino and gluino masses, respectively.  To leading order in the
gauge couplings, $C\simeq0.5$~\cite{Bolz:2000fu, Pradler:2006qh,
Buchmuller:2008vw}.\footnote{Note that $C$ has ${\cal O}(1)$ uncertainty due
to unknown higher order contributions and nonperturbative
effects~\cite{Bolz:2000fu}.} Remarkably, the observed dark matter relic
density $\Omega_{\text{DM}} h^2 = 0.11$~\cite{Komatsu:2010fb} is obtained for
typical supersymmetric parameters, $m_{3/2} \sim 100\,\text{GeV}$ and
$m_{\tilde g} \sim 1\,\text{TeV}$, and for reheating temperatures $T_R \sim
10^{10}\,\text{TeV}$ as required for thermal leptogenesis to generate the
observed baryon asymmetry~\cite{Fukugita:1986hr, Buchmuller:2004nz}. \medskip

However, the above scenario is not free of problems: If $R$-parity is
conserved, the decay of the next-to-lightest supersymmetric particle (NLSP)
into gravitino and standard model particles is suppressed by the Planck scale.
As a consequence, the NLSP becomes very long-lived and can dramatically affect
the successful prediction of the standard big bang nucleosynthesis (BBN)
scenario~\cite{Ellis:1984er, Sarkar:1995dd, Kawasaki:2004qu, Pospelov:2006sc,
Hamaguchi:2007mp, Pospelov:2008ta, Kawasaki:2008qe}.  Several alternative
scenarios have been proposed to consistently accommodate thermal leptogenesis,
gravitino dark matter and BBN, like assuming sneutrino~\cite{Kanzaki:2006hm}
or stop~\cite{DiazCruz:2007fc} NLSPs, diluting the NLSP abundance by entropy
production~\cite{Pradler:2006hh}, or introducing new NLSP decay channels into
hidden sector states~\cite{DeSimone:2010tr, Cheung:2010qf}.  Also, the
reheating temperature can be considerably reduced by considering non-thermal
leptogenesis scenarios, as for example recently proposed in mechanisms where
the universe is reheated through the decay of non-relativistic right-handed
neutrinos produced in false vacuum decay~\cite{Buchmuller:2010yy,
Buchmuller:2011soon}. 

Alternatively, it has been proposed that $R$-parity could be mildly
violated~\cite{Buchmuller:2007ui}.  Such a violation could be \fex produced by
dynamical breaking of U(1)$_{B-L}$ at low scales in the hidden
sector~\cite{Schmidt:2010yu}. It would lead to a rapid decay of the NLSP
before the onset of BBN, thus rendering its impact on the standard
cosmological picture negligible.  Notably, even though the gravitino is not
stable anymore in such a case, it still constitutes a viable dark matter
candidate because its decay into standard model particles is doubly suppressed
by the small $R$-parity breaking parameter as well as by the Planck
mass~\cite{Takayama:2000uz}, leading to lifetimes longer than the age of the
Universe.  \medskip

This $R$-parity violating scenario is particularly attractive because it opens
up the possibility to probe gravitino dark matter by searching for its decay
products in the cosmic-ray fluxes.  In addition to cosmic
anti-matter~\cite{Ishiwata:2008cu, Ibarra:2008qg} and
neutrino~\cite{Covi:2008jy} fluxes, the produced gamma-ray
flux~\cite{Takayama:2000uz, Buchmuller:2007ui, Lola:2007rw, Bertone:2007aw,
Ishiwata:2008cu,Ibarra:2007wg} is of high interest because, besides the
continuous component generated by the fragmentation of the Higgs and gauge
bosons, the gamma-ray energy spectrum typically features an intense line,
arising from the $\psi_{3/2} \rightarrow \gamma \nu$ two-body decay already at
tree level.\medskip

In the rest of this section, we will apply the above Fermi LAT limits on
gamma-ray lines to the decaying gravitino scenario.  First, we will summarize
the relevant aspects of bilinear $R$-parity violation, which will form the
theoretical framework of our analysis.  Second, we will present limits on the
size of $R$-parity violation, and discuss prospect for seeing long-lived
neutralino and stau NLSPs at the LHC, extending the analysis made
in~\cite{Bobrovskyi:2010ps}.\medskip

\subsection{\boldmath $R$-parity breaking model}
The supersymmetric standard model with bilinear $R$-parity breaking is, in
addition to the standard SU(3)$_c\times$SU(2)$_L\times$U(1)$_Y$ gauge
interactions, defined by mass-mixing terms between lepton and Higgs fields in
the superpotential 
\begin{align}
  W = W_{\text{MSSM}} + \mu_i H_u l_i\,,
\end{align}
as well as by the mass-mixing terms in the scalar potential as induced by
supersymmetry breaking
\begin{equation}
  {\cal L} = {\cal L}_{\text{soft}}^{\text{MSSM}} +  B_i H_u \tilde l_i +
  m_{id}^2\,\tilde l_i^\dagger H_d + \text{h.c.}\,.
\end{equation}
Here, $W_{\text{MSSM}}$ and ${\cal L}_{\text{soft}}^{\text{MSSM}}$ are the
usual $R$-parity conserving MSSM superpotential and scalar Lagrangian,
$H_{u/d}$ are the up/down-type Higgs doublets, $l_i$ the lepton doublets, and
$\mu_i$, $B_i$ and $m_{id}^2$ are the $R$-parity violating couplings.  As
proposed in Ref.~\cite{Bobrovskyi:2010ps}, it is convenient to work in a
lepton-higgs basis where the mass mixings $\mu_i$, $B_i$ and $m_{id}^2$ are
traded for $R$-parity breaking Yukawa couplings.  This can be achieved by a
supersymmetric rotation of the superfields, followed by a non-supersymmetric
rotation of the scalar fields alone. Besides the $R$-parity breaking Yukawa
couplings, which will become relevant for the decay of the stau NLSP below,
these rotations also generate mass-mixing terms between the neutralinos and
the neutrinos. This neutrino-neutralino mixing, which can be parameterized by
the dimensionless parameter $\zeta$, finally induces the gravitino decay mode
$\psi_{3/2}\to\gamma\nu$ as well as the decay of neutralino NLSPs (see
Ref.~\cite{Bobrovskyi:2010ps} for a definition of $\zeta$ in terms of the
bilinear $R$-parity violating couplings $\mu_i$, $B_i$ and $m_{id}^2$).
\medskip

We will consider two typical sets of boundary conditions for the supersymmetry
breaking parameters of the MSSM at the grand unification (GUT) scale for
definiteness. The first one corresponds to equal scalar and gaugino masses
\begin{align}
  \text{(A)}\qquad  m_0 = m_{1/2},\quad a_0=0, \quad \tan \beta =10\,,
  \label{eq:boundaryA} 
\end{align}
for which the bino-like neutralino $\chi_1^0$ is the NLSP. In the second one,
which corresponds to no-scale models or gaugino mediation, 
\begin{align}
  \text{(B)} \qquad m_0 = 0, \quad  m_{1/2},\quad a_0=0, \quad \tan
  \beta =10\,,\label{eq:boundaryB} 
\end{align}
the lightest stau $\tilde \tau_1$ is the NLSP. In both cases, $\tan \beta =
10$ has been chosen as a representative value, and the trilinear scalar
coupling $a_0$ has been set to zero for simplicity. The universal gaugino mass
$m_{1/2}$ remains as the only independent variable. For both sets of boundary
conditions, the gaugino masses $M_{1,2,3}$ satisfy the following relations at
the electroweak scale 
\begin{align}
  \frac{M_3}{M_1} \simeq 5.9\,,\qquad   \frac{M_2}{M_1} \simeq 1.9\,.
  \label{eq:gauginomasses}
\end{align}
\medskip

Electroweak precision tests (EWPT) yield important lower bounds on the
superparticle mass spectrum~\cite{Buchmuller:2008vw}.  For a neutralino NLSP,
the most stringent constraint comes from the Higgs potential.  The universal
gaugino mass $m_{1/2}$ is required to be high enough in order for the Higgs
mass to fulfills the LEP lower bound $m_h >
114.4$\,GeV~\cite{Nakamura:2010zzi}. This implies the lower limit $\mN \gtrsim
130$\,GeV.\footnote{Note that $m_{\chi^0_1}\simeq M_1$ with good
accuracy~\cite{Bobrovskyi:2010ps}.} However, allowing negative $a_0$ or scalar
masses much larger than $m_{1/2}$ at the GUT scale would weaken this limit,
and we will take $m_{\chi^0_1}>100\GeV$ as a lower bound for the neutralino
mass subsequently.  In the stau NLSP case, the lower bound comes from the
absence of pair production of heavy charged particles at LEP and reads
$m_{\tilde\tau_1}>100$\,GeV~\cite{Nakamura:2010zzi}. 

For a typical effective neutrino mass $\widetilde m_1 =10^{-3}$\,eV,
successful thermal leptogenesis requires a minimal reheating temperature of
$T_R \sim 10^9$\,GeV~\cite{Buchmuller:2004nz}.  Using
Eq.\,\eqref{eq:gravitinoabundance} together with a lower bound on the gluino
mass $m_{\tilde g} \gtrsim 500$\,GeV~\cite{Alwall:2008ve}, this implies a
lower bound for the gravitino mass $m_{3/2} \gtrsim 5$\,GeV.

In addition to the lower limits, NLSP mass upper limits follow from the
requirement that the gravitino does not overclose the Universe. Indeed by
rewriting Eq.~\eqref{eq:gravitinoabundance} one obtains the constraint
\begin{align}
  m_{\text{NLSP}} \simeq 310\,\text{GeV} \left( \frac{\xi}{0.2} \right)
  \bigg( \frac{m_{3/2}}{100\,\text{GeV}} \bigg)^{1/2} \left(
  \frac{10^9\,\text{GeV}}{T_R} \right)^{1/2}\;,
  \label{eq:NLSPupbound}
\end{align}
where $\xi \equiv m_{\text{NLSP}}/m_{\tilde g}$ is implicitly fixed by the
supersymmetry breaking boundary conditions~\cite{Buchmuller:2008vw}.  Since
$m_{\text{NLSP}} \propto m_{3/2}^{1/2}$, requiring the gravitino to be the LSP
induces absolute upper bounds on the NLSP masses.  In the case of the
neutralino NLSP, Eq.~\eqref{eq:NLSPupbound} implies $\mN \lesssim 690$\,GeV
for $\xi = 1/5.9$, and is essentially independent of $m_0$ and $\tan \beta$.
For the stau NLSP, $\tan \beta = 10$ yields $\xi = 1/6.2$, which consequently
leads to the more stringent bounds $\mstau \lesssim 615$\,GeV.  Note that
there is a strong dependence on $\tan \beta$ in that
case~\cite{Buchmuller:2008vw}, and that $\xi$ decreases with increasing $\tan
\beta$.

\subsection{Fermi LAT constraints on the NLSP decay length}
\label{sec:zetaconstraint}
The gravitino decay mode relevant for the discussion in this paper is the
decay into photons and neutrinos via $\psi_{3/2}\to\gamma\nu$. However, decays
into $W^\pm\ell^\pm$, $Z^0\nu$ and $h^0\nu$ final states are also possible,
and have often large branching ratios. Below the corresponding kinematic
thresholds, three-body decay with intermediate massive gauge bosons can become
important in some cases~\cite{Choi:2010xn}.

Additionally to the gamma-ray line from $\psi_{3/2}\to\gamma \nu$, the
different decay modes will produce a gamma-ray continuum, coming from Higgs
and gauge boson fragmentation as well as from final state radiation of the
charged leptons. This continuous spectrum has a cutoff at the position of the
gamma-ray line, namely at half of the gravitino mass. Since we required in the
above line search that all gamma-ray fluxes except the line follow a power-law
locally around the line's energy (\textit{cf.}~section~\ref{sec:methods}), a
too large continuum contribution from gravitino decay would render our line
search strategy invalid.  We expect this to become relevant for branching
ratios into gamma-ray lines below $10^{-3}\dots10^{-2}$, and postpone a
detailed analysis of this effect to future work.  Using the branching ratios
presented in Refs.~\cite{Ishiwata:2008cu, Covi:2008jy}, we find that this
problem can become severe for gravitino masses above a few hundred GeV,
$m_{3/2}\gtrsim300\GeV$. However, one has to keep in mind that the exact
branching ratio into lines is in principle model-dependent. In what follows we
assume for simplicity that no continuum contribution disturbs our line search,
and we present limits over the full range of accessible gravitino masses.
\medskip

\begin{figure}[t]
  \begin{center}
    \includegraphics[width=0.7\linewidth]{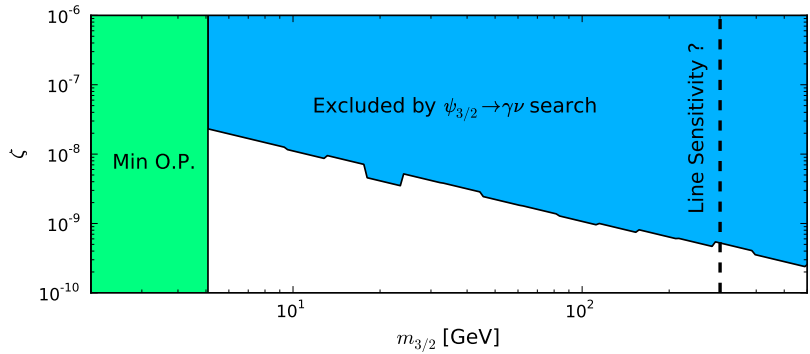}
  \end{center}
  \caption{Upper bounds on the $R$-parity violation parameter $\zeta$, derived
  from the Fermi LAT gamma-ray line limits in Fig.~\ref{fig:bounds_DDM}. For
  thermal leptogenesis, overproduction (O.P.) of gravitinos already excludes
  the left green region. As discussed in the text, the adopted strategy for
  line searches might fail for gravitino masses above a few hundred GeV
  (dashed black line).} \label{fig:bounds_zeta}
\end{figure}

The gravitino inverse decay rate into photon/neutrino pairs is given
by\footnote{The correction factor is less than 10\% for bino masses under
consideration~\cite{Bobrovskyi:2010ps}.}~\cite{Takayama:2000uz,
Bobrovskyi:2010ps} 
\begin{align}
  \label{eq:Gdecayrate}
  \Gamma^{-1}_{\psi_{3/2}\to\gamma \nu} = \frac{32 \sqrt{2}}{\alpha \zeta^2}
  \frac{\GF
  \MP^2}{\mG^3} \frac{M_1^2 M_2^2}{(M_2-M_1)^2}\left( 1+ {\cal O}\left( s_{2
  \beta} \,
  \frac{m_Z^2}{\mu^2} \right) \right)\,,
\end{align} 
where $\alpha$ is the electromagnetic fine structure constant, $\MP =
2.4\times 10^{18}$\,GeV the reduced Planck mass, and $\GF = 1.16\times
10^{-5}$\,GeV$^{-2}$ is the Fermi constant.  Using this expression, the limits
on the gamma-ray line flux can be translated into upper limits on the
$R$-parity breaking parameter $\zeta$. A conservative bound on $\zeta$ is
obtained for a given gravitino mass when using the corresponding maximally
allowed bino mass. The latter is obtained by combining
Eq.\,\eqref{eq:gravitinoabundance} and \eqref{eq:gauginomasses} considering
the lowest reheating temperatures allowed in the thermal leptogenesis
scenario, \textit{i.e.} $T_R \sim 10^9$\,GeV. The limits on the $R$-parity
violation parameter $\zeta$ are presented in Fig.~\ref{fig:bounds_zeta}.  As
discussed above, for gravitino masses above a few hundred GeV, our line search
may fail due to a too large gamma-ray continuum contribution, which is
indicated by the dashed line.  Note also that, at these high gravitino masses,
the production of anti-protons in gauge boson fragmentation could further
constrain the $\zeta$ parameter~\cite{Ishiwata:2008cu, Ibarra:2008qg}.  The
implications of our gamma-ray line limits on the decay of the NLSP are
discussed next.

\subsubsection{Stau NLSP}
In the case of the $\tilde \tau_1$-NLSP, the total decay width of the lightest
mass eigenstate is a mixture of left and right handed partial decays
\begin{align}
  \Gamma_{\tilde \tau_1} = \sin^2\theta_{\tilde \tau} \,\Gamma_{\tilde \tau_L}
  + \cos^2\theta_{\tilde \tau} \,\Gamma_{\tilde \tau_R}\,.
  \label{eq:ctaustau}
\end{align}
Since the $R$-parity breaking Yukawa couplings are typically proportional to
the ordinary Yukawa couplings, decays into second and third families dominate.
For definiteness, we will below assume a flavor structure as described in
Ref.~\cite{Bobrovskyi:2010ps}, which is based on a Froggatt-Nielsen U(1)
flavor symmetry.  In such a framework, the chiral state decays are dominated
by the following channels~\cite{Bobrovskyi:2010ps}
\begin{subequations}
  \begin{align}
    \tilde \tau_R &\rightarrow \tau_L \nu, \mu_L \nu\,,\\
    \tilde \tau_L & \rightarrow \bar t_R b_L\,.
  \end{align}
\end{subequations}
The corresponding decay rates are
\begin{subequations}
  \begin{align}
    \Gamma_{\tilde \tau_L}&= \frac{\epsilon^2}{16 \pi v^2} \frac{m_t^2
    }{\mstau^3} \,3 \, (\mstau^2
    - m_t^2 - m_b^2)  \sqrt{[\mstau^2 - (m_t + m_b)^2] [\mstau^2 - (m_t -
    m_b)^2]}\,,\\
    \Gamma_{\tilde \tau_R}&= \frac{\epsilon^2}{16 \pi v^2} 
    \,\frac{m_t^2 }{\mstau^3} \,(\tan \beta)^2 \,(\mstau^2
    - m_\tau^2)^2 \,,
  \end{align}
\end{subequations}
where the dimensionless parameter $\epsilon$ is directly related to the
$R$-parity violating Yukawa couplings (see Ref.~\cite{Bobrovskyi:2010ps} for
details).  In principle $\zeta$ and $\epsilon$ are independent parameters
because they stem from different linear combinations of the bilinear
$R$-parity violating couplings $\mu_i$, $B_i$ and $m_{id}^2$. We choose
$\zeta\simeq\epsilon$ in order to set limits on the stau decay length.
However, one has to keep in mind that in principle $\zeta$ could be set to
values much smaller than $\epsilon$ by a proper choice of the parameters
$\mu_i$, $B_i$ and $m_{id}^2$.  The behavior of the mixing angle
$\theta_{\tilde \tau}$ with the stau mass $\mstau$ can be deduced from the
RGEs using the boundary conditions Eq.~\eqref{eq:boundaryB}.

\begin{figure}[t]
  \begin{center}
    \includegraphics[width=0.80\linewidth]{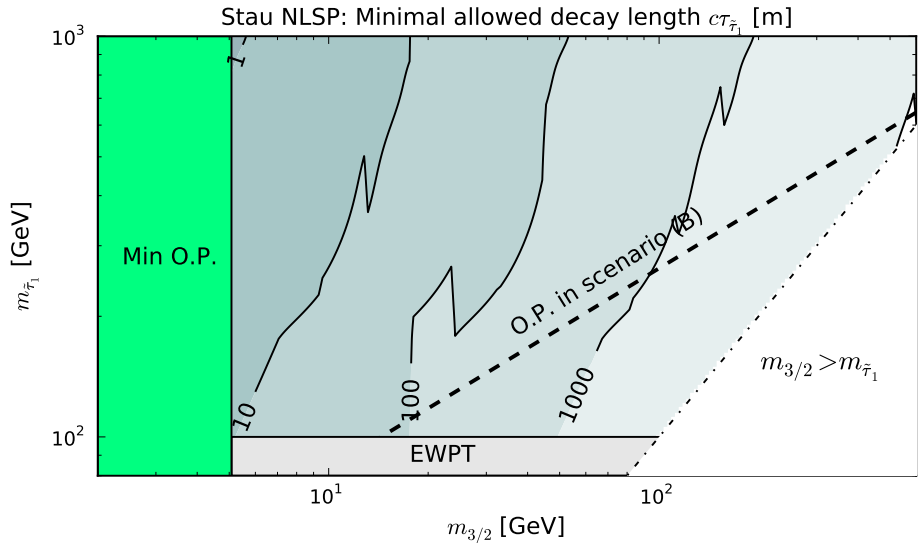}
  \end{center}
  \caption{Contour plot of lower bounds on the stau NLSP decay length coming
  from the gamma-ray line constraints on the gravitino lifetime, as function
  of the stau and gravitino masses, $m_{\tilde\tau_1}$ and $m_{3/2}$
  respectively. The lower gray region is excluded by electroweak precision
  tests (EWPT).  For thermal leptogenesis, overproduction (O.P.) of gravitinos
  excludes at minimum the left green region, a limit which strengthens to the
  black-dashed line when assuming the universal boundary conditions (B),
  \textit{cf.}~Eq.~\eqref{eq:boundaryB}. The lower-right exclusion comes from
  the gravitino LSP requirement. Note that for high gravitino masses above a
  few hundred GeV our adopted line search strategy might overestimate the
  limits on the NLSP decay length, because the branching ratio into gamma-ray
  lines can become very  small (see text).}
  \label{fig:Staudecaylength}
\end{figure}

Using the upper limits on the $R$-parity breaking parameter $\zeta$ from
Fig.~\ref{fig:bounds_zeta}, we can derive lower bounds on the stau decay
length.  Our results are shown in Fig.~\ref{fig:Staudecaylength}.  The
parameter space is already constrained by EWPT and overproduction bounds, and
the lower limits on the neutralino decay length vary between 100\,m and
10\,km.  It is interesting that if such particle were to be produced at the
LHC, a sizable amount of their decay could take place in the
detector~\cite{Ishiwata:2008tp}.  We obtain a lowest possible decay length
$c\tau_{\tilde \tau} \simeq 85$\,m  for $\mG \simeq 16$\,GeV and $\mstau
\simeq 100$\,GeV. 

\subsubsection{Neutralino NLSP}
A neutralino NLSP heavier than 100\,GeV dominantly decays into $W^\pm\ell^\mp$
and $Z^0\nu$~\cite{Mukhopadhyaya:1998xj, Chun:1998ub}.  The corresponding
decay width is directly proportional to the $R$-parity breaking parameter
$\zeta$ squared, which also enters the gravitino decay width in
Eq.~\eqref{eq:Gdecayrate}.  As a consequence, the two quantities can be
related through~\cite{Bobrovskyi:2010ps}
\begin{align}
  \tau_{\chi_1^0} = \frac{c_w^2}{2 \sqrt{2}} \frac{(M_2-M_1)^2}{M_2^2}
  \frac{\mG^3}{\GF \MP^2 \mN^3} \,\frac{\Gamma_{\psi_{3/2}\to\gamma\nu}^{-1}}{
  2f(\mN,m_W) + f(\mN,m_Z)}\,,
  \label{eq:ctauneutralino}
\end{align}
where the phase space factor $f$ is defined by
\begin{align}
  f(m_1,m_2) = \left(1- \frac{m_2^2}{m_1^2} \right)^2 \left( 1+
  2\,\frac{m_2^2}{m_1^2}\right)\,.
\end{align}

\begin{figure}[t]
  \begin{center}
    \includegraphics[width=0.80\linewidth]{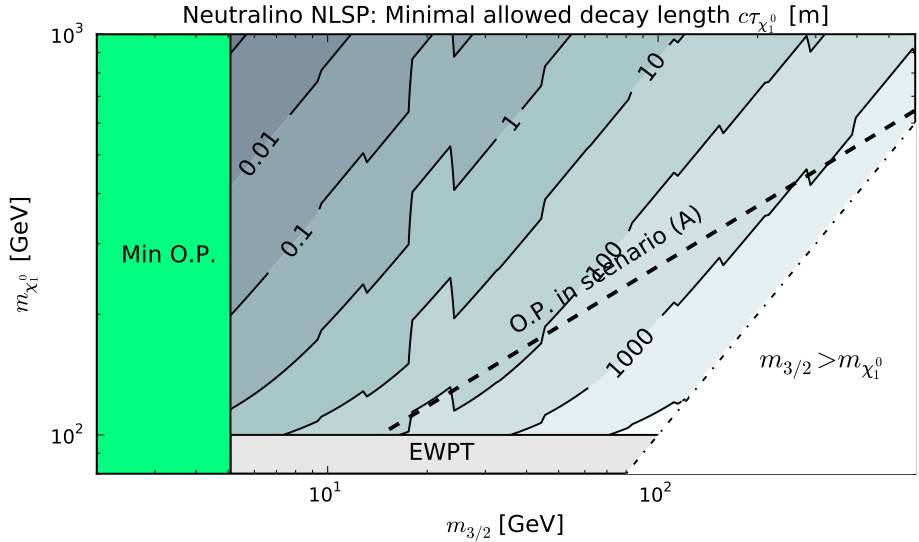}
  \end{center}
  \caption{Like Fig.~\ref{fig:Staudecaylength}, but for a neutralino NLSP.}
  \label{fig:Neutralinodecaylength}
\end{figure}

Using the gaugino mass relation from Eq.~\eqref{eq:gauginomasses}, lower
bounds on the neutralino decay length $c\tau_{\chi_1^0}$ can then be derived
from Tab.~\ref{tab:bounds} and Fig.~\ref{fig:bounds_DDM}. Our results are
summarized in Fig.~\ref{fig:Neutralinodecaylength}.  For the parameter space
allowed by EWPT and overproduction bounds, we obtain minimal decay lengths
${\cal O}(100\,\text{m}-10$\,km), which are in the range of detectability of
the LHC~\cite{Ishiwata:2008tp}.  Decay lengths as small as $c\tau_{\chi_1^0}
\simeq 50$\,m are allowed for $\mG \simeq 17$\,GeV et $\mN \simeq 110$\,GeV.

\section{Conclusions}
We have presented a dedicated search for gamma-ray lines with energies between
1 and 300 GeV in the diffuse gamma-ray fluxes measured by Fermi LAT between 4
Aug 2008 and 17 Nov 2010, based on the public `DataClean' event class. Such
lines could be produced in dark matter annihilation or decay, and they are
highly relevant for dark matter searches since they provide an essentially
background free signature. To calculate the significance of potential
gamma-ray lines in the diffuse flux, we used a binned profile likelihood
method and fitted the observed energy spectrum locally with a line signal as
expected from decaying or annihilating dark matter inside the Galactic halo
plus a power law that resembles the astrophysical background fluxes. No
gamma-ray line was found at $5\sigma$~confidence level, and we derived
conservative upper limits on the gamma-ray line flux that comes from regions
with high Galactic latitudes or from the Galactic center
(Tab.~\ref{tab:bounds}). We also discussed the apparent presence of several
line-like features at the few percent level in the public data, which appear
at energies below 20 GeV with $\sim3\sigma$ significance and which are most
probably due to instrumental effects.  

\sloppy
Our conservative upper limits on the partial annihilation cross section of
WIMP dark matter into $\gamma$-pairs, $\langle\sigma
v\rangle_{\psi\psi\to\gamma\gamma}$, improve the EGRET limits in the range
$E_\gamma = 1$--$10\GeV$ by up to an order of magnitude, while the previous
Fermi LAT results ($E_\gamma = 30$--$200\GeV$) are updated with two years of
data and extended to the larger energy range $E_\gamma=1$--$300\GeV$
(Fig.~\ref{fig:bounds_ADM}). Despite the increased statistics and the choice
of a sky-region with optimized signal-to-noise ratio, the previous Fermi LAT
limits are however only slightly improved. The reason is a somewhat worse
energy resolution of the public data with respect to the dedicated Fermi LAT
analysis as well as our strategy to always quote conservative limits in
certain energy bands. The lower bounds on the inverse decay widths of dark
matter into monochromatic photons, $\Gamma_{\psi\to\gamma\nu}^{-1}$, are at
the level of $6\times10^{28}\s$ over most of the considered energy regime
(Fig.~\ref{fig:bounds_DDM}). Our results also constrain annihilation and decay
channels with $\gamma Z^0$ final states.  \smallskip

\fussy

In a $R$-parity breaking supersymmetric framework with gravitino LSP, we used
the results of our gamma-ray line search to derive lower bounds on the
gravitino lifetime. Considering supergravity models with two different types
of universal boundary conditions at the grand unification scale, these
lifetime constraints were used to set lower limits on the corresponding NLSP
decay lengths. For gravitino and NLSP masses compatible with electroweak
precision tests and overproduction constraints, we obtain ${\cal
O}(100\,\text{m}-10\,\text{km})$ lower bounds on the NLSPs decay lengths.
Interestingly, all these decay lengths would be accessible at the LHC.

\paragraph{Note.}
During the final stages of this work, first LHC limits on the CMSSM parameter
space were published by the CMS collaboration~\cite{Collaboration:2011tk},
which will further constrain the superparticle masses discussed in this paper.
Furthermore, the anonymous referee pointed us to a talk about the ongoing
gamma-ray line analysis by the Fermi LAT collaboration, where preliminary
results are presented that are, where they overlap, in agreement with our
findings, and where line-like artefacts at lower energies in the current
public data are mentioned.\footnote{See talk by E.~Bloom at Aspen Winter
Workshop on Indirect and Direct Detection of Dark Matter, Feb 2011,
\url{http://www.slac.stanford.edu/exp/glast/aspen11/talks.asp}}

\section*{Acknowledgements}
We thank Sergei Bobrovskyi, Wilfried Buchm\"uller, Jan Hajer and Alejandro
Ibarra for useful discussions, David Paneque and Jeremy S.~Perkins for helpful
information on the Fermi LAT data, and Anthony R.~Pullen for information about
the EGRET analysis.

\bibliographystyle{JHEP}
\bibliography{}

\end{document}